\documentclass[prl,twocolumn,superscriptaddress,showpacs,pdfoutput=1]{revtex4-1}
\usepackage{times}
\usepackage{graphicx}
\usepackage{epsfig}
\usepackage{amssymb,amsmath,amsfonts,hyperref}
\usepackage{wasysym}
\usepackage{bm}
\usepackage{tikz}
\usepackage{latexsym}
\usepackage{eucal}
\usepackage{float}
\usepackage{verbatim}
\usepackage[normalem]{ulem}
\usepackage{epstopdf}
\usepackage{color}
\usepackage{braket}

\usepackage{youngtab}
\usepackage{ifthen}
\usepackage{xspace}
\usepackage{multirow}
\usepackage{dsfont}

\newcommand{\be}{\begin{equation}}
\newcommand{\ee}{\end{equation}}
\newcommand{\mN}{{{\mathbb{N}}}}
\newcommand{\bc}{{{\bar{c}}}}
\newcommand{\cp}{{{c^{\prime}}}}
\newcommand{\bcp}{{{\bar{c^{\prime}}}}}

\newcommand{\eea}{\end{eqnarray}}

\newcommand{\su}[3]{
        \ifthenelse{\equal{#2}{0}}{\ifthenelse{\equal{#1}{0}}{}{\mult{#1}}\overset{\bf #3}{\underset{\phantom{.}}{\bullet}}}{\ifthenelse{\equal{#1}{0}}{}{\mult{#1}}\underset{\phantom{.}}{\overset{\bf #3}{\Yboxdim9pt\yng(#2)}}}
}
\newcommand{\mult}[1]{{ \fcolorbox{gray!70}{gray!70}{\textcolor{white}{\footnotesize \bf #1}}}\;}

\begin{document}

\title{Weakly first-order quantum phase transition between Spin Nematic and Valence Bond Crystal Order in a square lattice SU(4) fermionic model}

\author{Pranay Patil}
\affiliation{Laboratoire de Physique Th\'eorique, Universit\'e de Toulouse, CNRS, UPS, France}

\author{Fabien Alet}
\affiliation{Laboratoire de Physique Th\'eorique, Universit\'e de Toulouse, CNRS, UPS, France}

\author{Sylvain Capponi}
\affiliation{Laboratoire de Physique Th\'eorique, Universit\'e de Toulouse, CNRS, UPS, France}

\author{Matthieu Mambrini}
\affiliation{Laboratoire de Physique Th\'eorique, Universit\'e de Toulouse, CNRS, UPS, France}

\begin{abstract}
We consider a model Hamiltonian with two SU$(4)$ fermions per site on a square lattice, showing a competition between bilinear and biquadratic interactions. This model has generated interest due
to possible realizations in ultracold atom experiments and existence of spin liquid ground states.
Using a basis transformation, we show that part of the phase diagram is amenable to quantum Monte Carlo simulations without a sign problem.
We find evidence for spin nematic and valence bond crystalline phases, which are separated by a weak first order phase transition. A U($1$) symmetry is found to emerge in the valence bond crystal histograms, suggesting proximity to a deconfined quantum critical point.  Our results are obtained with the help  of a loop algorithm which allows large-scale simulations of bilinear-biquadratic SO($N$) models on arbitrary lattices in a certain parameter regime.
\end{abstract}

\maketitle

{\it Introduction --}
Extended symmetries often offer a way to realize new phases of matter in simple models of strongly correlated quantum systems. An  important motivation for extended symmetries comes from studying the limit where the number of internal degrees of freedom $N$ becomes large, an ubiquitous tool in theoretical physics~\cite{Stanley68,hooft,moshe_largeN}. Indeed this large-$N$ limit is often tractable analytically, allowing a better physical understanding and giving a starting point for an expansion aimed to characterize the small-$N$, physical, cases. In quantum magnetism, this approach was pionereed by enlarging the symmetry group to SU$(N)$ where it was for instance predicted, using field-theoretical analysis~\cite{read_features_1989,read_spin-peierls_1990}, that the well-known antiferromagnetic (N\'eel) ordered phase present on the square lattice at small $N$ is replaced by a valence-bond crystal (VBC) that breaks lattice symmetries at large $N$. For several SU$(N)$ representations and different lattices, numerical studies have confirmed the existence of ground-states without magnetic long-range order~\cite{harada2003neel,corboz_simultaneous_2011,Corboz2012prx,Corboz2012,Corboz2013,Nataf2016}. Extended symmetries are not only useful as a theoretical knob, but are also meaningful to describe experimental systems: for instance,  SU$(4)$ symmetry is relevant for materials with strong spin-orbit coupling~\cite{Kugel1982,SU4_Kitaev} while SO$(4)$ symmetry has been suggested for twisted bilayer graphene~\cite{you_superconductivity_2019}. In atomic physics, alkaline-earth ultracold atoms show an almost perfect realization of SU$(N)$ symmetry groups with high values of $N$~\cite{cazalilla_ultracold_2014,Gorshkov2010,pagano_one-dimensional_2014,deSalvo2010,Tey2010,Taie2010} while spin-3/2 fermions can realize SO(5) symmetry~\cite{Wu2003,Wu2006}. Recent experiments with ultracold atomic systems show that low temperatures can be reached for SU(N)-symmetric alkaline-earth elements~\cite{Sonderhouse2020} while a filling of two fermions per site can be realized~\cite{Hartke2022} as it avoids three-body losses.

The competition between different energy terms, compatible with extended symmetries, is another fruitful approach to engineer unconventional phenomena~\cite{designer}. For instance, the competition between VBC and N\'eel ordered phases found in large-$N$ theories triggered a large interest due to the possibility of a generically continuous deconfined quantum critical point (DQCP)~\cite{Senthil2004,Senthil2004prb,sandvik2007evidence} between these two phases of matter, in contradiction with naive expectations from Landau-Ginzburg theory. A continuous transition can be observed numerically by either artificially treating $N$ as a continuous parameter~\cite{Beach2009}, or due to the competition between terms involving two and four or more spins, for a large variety of SU$(2)$ and SU$(N)$ models \cite{kaul2011quantum,harada2013possibility,kaul2012lattice,sandvik2007evidence,sandvik2010continuous,lou2009antiferromagnetic}. An excellent agreement with large-$N$ DQCP predictions is obtained as $N$ is increased~\cite{kaul2012lattice}. For magnetic systems hosting spins larger than $1/2$, another important competing term compatible with SU$(N\geq 2)$ symmetry is a biquadratic coupling between two spins. Biquadratic terms are also relevant for cold-atomic systems~\cite{Yip2003,Imambekov2003,Eckert_2007,Brennen_2007,puetter2008theory}. For spin-1 systems in two dimensions (2D), it is possible to obtain a (spin) nematic (or ferroquadrupolar) ground-state that breaks SU$(2)$ symmetry, without any local magnetization, but with a finite quadrupolar order~\cite{nematic_review}. For instance, the bilinear-biquadratic Heisenberg model on the square lattice exhibits a very rich phase diagram~\cite{Papanicolaou1988,Toth2012,Niesen2017}, including a nematic phase. For a quasi-one-dimensional spin-1 model, Harada {\it et al.}~\cite{harada2006dimer} found numerical evidence for a continuous transition between a nematic and a VBC phase. The VBC phase does not survive to the isotropic 2D limit, leading instead to a magnetically ordered phase which exhibits a first-order transition to the nematic phase. This system was analyzed with a bond-operator treatment in Ref.~\cite{puetter2008theory}, predicting a generic first-order nematic-VBC transition, along with a discussion of possible spin liquid behavior for SO$(N)$ symmetry at large $N$. On the other hand, a general discussion of nematic behavior from the perspective of a continuum field theory incorporating the role of Berry phases~\cite{grover2007quantum} allows for a continuous DQCP to a VBC phase for quasi-one-dimensional SO$(3)$ models. In a subsequent quantum Monte Carlo (QMC) numerical study, Kaul~\cite{kaul2012spin} showed that a pure biquadratic model on a triangular lattice, which is known to host a nematic ground-state and has an extended SO$(3)$ symmetry~\cite{Lauchli2006,kaul2012spin}, can exhibit VBC or spin-liquid ground-states when the symmetry is extended to SO$(N)$ for large-enough $N$ and/or in presence of further competing interactions~\cite{kaul2015spin}. The phase transitions between spin nematic and VBC phases were found to be  discontinuous.

In this work, we consider a square lattice model built out of two SU$(4)$ fermions per site, showing a competition between bilinear and biquadratic terms. This model has been discussed earlier~\cite{Marston89,affleck_su2n_1991,paramekanti_sun_2007,gauthe2020quantum,kim2019dimensional,wang2014competing} with predictions of a rich phase diagram with N\'eel order, VBC, ferromagnet and charge-conjugation symmetry broken phases~\cite{paramekanti_sun_2007}, as well as of critical spin liquids phases from a projected entangled pair states (PEPS) ansatz~\cite{gauthe2020quantum}. We use an {\it exact} mapping to an SO$(n_c)$ model with $n_c=6$ colors, and show that part of the phase diagram can be simulated exactly using QMC with no sign problem.
\begin{figure}[t]
\includegraphics[width=\hsize]{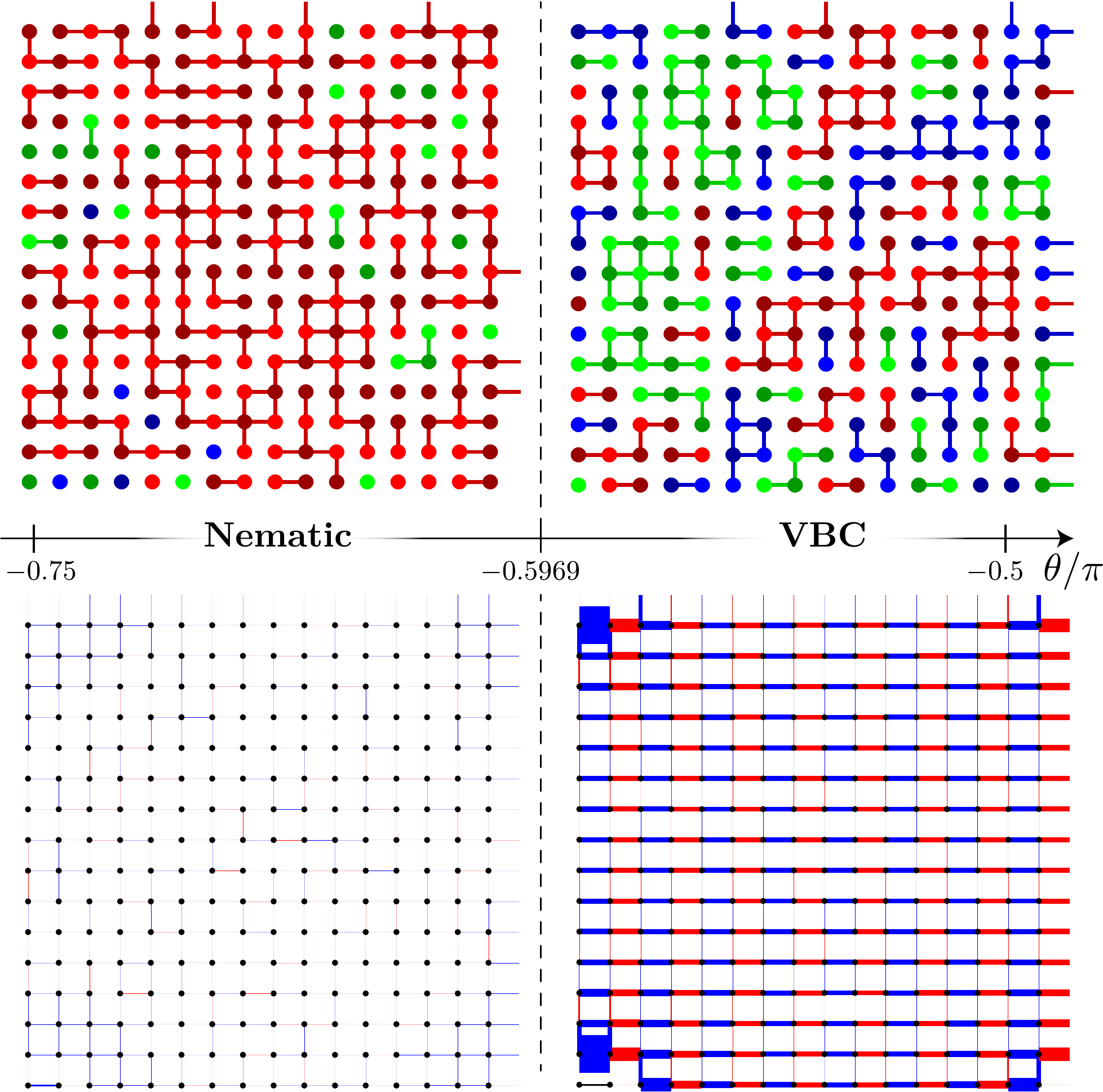}
\caption{\label{fig:pd} Phase diagram in the sign-free region
$\{\theta_{\rm SF}\}=[-3/4 \pi, - \pi/2]$: Nematic to VBC transition for
$\theta_c=-0.5969(1)\pi$. Upper panels : representative configurations (see text)
constructed from snapshots of Monte Carlo configurations. Bottom panels :  real-space pattern of the connected bond correlator $D_{\rm corr}(b)-D_{\rm corr}^{\rm avg}$ with respect to the bottom left bond
(shown in black). Blue and red mean positive and negative respectively and
the bond thickness denotes the magnitude. Data are presented for
$\theta=-0.74\pi$ (nematic) and $\theta=-0.5\pi$ (VBC).}
\label{phaseschematic}
\end{figure}

{\it Model definitions -- }
We first define the model with two SU$(4)$ fermions per lattice site, which form a 6-dimensional space at each site~\cite{Marston89,affleck_su2n_1991,paramekanti_sun_2007,gauthe2020quantum}, with the following Hamiltonian
\begin{equation}
{\cal H}= J \sum_{\langle ij \rangle } {\bf S}_i \cdot {\bf S}_j + \frac{K}{4} \sum_{\langle ij \rangle } ( {\bf S}_i \cdot {\bf S}_j )^2,
\end{equation}
where $J=\cos(\theta)$ and $K=\sin(\theta)$.  By analogy with the usual SU($2$) spin case, the 15-dimensional vector ${\bf S}$ is formed by the generators of SU($4$) and the ``spin'' interaction ${\bf S} \cdot {\bf S}$ can be expressed as a linear combination of symmetric projectors on different irreducible on-site representations (see Ref.~\cite{gauthe2020quantum}). The model exhibits an enlarged SU($6$) symmetries at $J=0$ ($\theta=\pm \pi/2$, with fundamental representation on one sublattice and conjugate on the other one) and $J=K$ ($\theta=-3/4 \pi$,\, $\pi/4$, with fundamental representation on each lattice site). QMC studies of the Hubbard model at large interaction find a critical or weakly ordered N\'eel phase~\cite{assaad_phase_2005,wang_competing_2014} at $\theta=0$. The Hamiltonian can be alternatively written in a basis with $n_c=6$ colors degree of freedom ($c=1\ldots 6$), encoding the six possible states on each site (see Sup. Mat.~\cite{supmat}). Denoting by $\bar{c}=n_c+1-c$ the complementary color of $c$, the Hamiltonian reads (up to an irrelevant constant):
\begin{equation}
\label{eqHcolor}
{\cal H} = \sum_{\langle i,j\rangle}   \sum_{c,c'} \left(J | cc' \rangle \langle c' c| + (K-J) | c\bar{c} \rangle \langle c' \bar{c'}| \right).
\end{equation}
In this form, the model has non-positive matrix elements when $J\leq 0$ and $K\leq J$, resulting in the sign-problem free region $\{\theta_{\rm SF}\}=[-3/4\pi,-\pi/2]$ for QMC simulations in this color basis. A variational wave-function analysis~\cite{paramekanti_sun_2007} predicts the existence of a VBC (dimerized) and ferromagnetic phases in this region. Quite interestingly, PEPS computations~\cite{gauthe2020quantum} find in the same region indications for a lack of ordering, and two variational (critical) spin liquids wave-functions with very competitive energies. We adapt (see details in Sup. Mat.\cite{supmat}) an efficient QMC loop algorithm for bilinear-biquadratic spin 1 models~\cite{kawashima2004}, to simulate the model Eq.~\ref{eqHcolor} in $\{\theta_{\rm SF}\}$. We perform simulations of square lattice samples with $N=L^2$ sites with linear size $L$ up to $96$, and up to inverse temperature $\beta=2L$ in units of $1/\sqrt{J^2+K^2}$ to reach ground-state properties. Our results can be summarized as follows (see Fig.~\ref{fig:pd}). We find that the region $\{\theta_{\rm SF}\}$ hosts two ordered phases: a VBC phase (known~\cite{harada2003neel} to exist at $\theta=-\pi/2$) as well as a nematic phase defined by a spontaneous symmetry-breaking choice of color pairs, which appears to have been missed earlier. Cartoon representations of QMC configuration snapshots for
these two phases are provided in Fig.~\ref{fig:pd}, where states $c$ and
$\bar{c}$, which form a nematic pair, are represented by different shades of
the same color, and bonds of the same color are drawn between neighboring
lattice sites hosting $c$ and $\bar{c}$. In the nematic phase, one
of three possible colors dominates, whereas in the VBC phase, there is no
dominance of a single color, but most neighboring lattice sites are
connected by bonds. The VBC pattern is not easily discernible and
a more detailed study of the dimer correlation in the VBC
phase is presented later in this manuscript.
We provide evidence for a very weak first-order transition between the VBC and the nematic phase at $\theta_c=-0.5969(1)\pi$. The VBC phase is furthermore found to exhibit an emergent U$(1)$ behavior all along the range $[\theta_c,-\pi/2]$ amenable to QMC, restricting our ability to classify this phase into columnar, plaquette or mixed order \cite{ralko2008generic,yan2021widely}. This emerging symmetry is strongly reminiscent of the behavior observed at or close to a DQCP \cite{sandvik2007evidence,Jiang_2008,lou2009antiferromagnetic,VBS-Sandvik,nahum_deconfined,pujari2013neel,sreejith2019emergent}. We suggest that our results could correspond to a runaway flow close to a potential DQCP fixed point, similar to the theory between nematic and VBC phases presented by Grover and Senthil~\cite{grover2007quantum} for an SO$(3)$ quasi-one-dimensional model, calling for a similar analysis for the SO$(6)$ case.

\begin{figure}[t]
\includegraphics[width=0.9\linewidth]{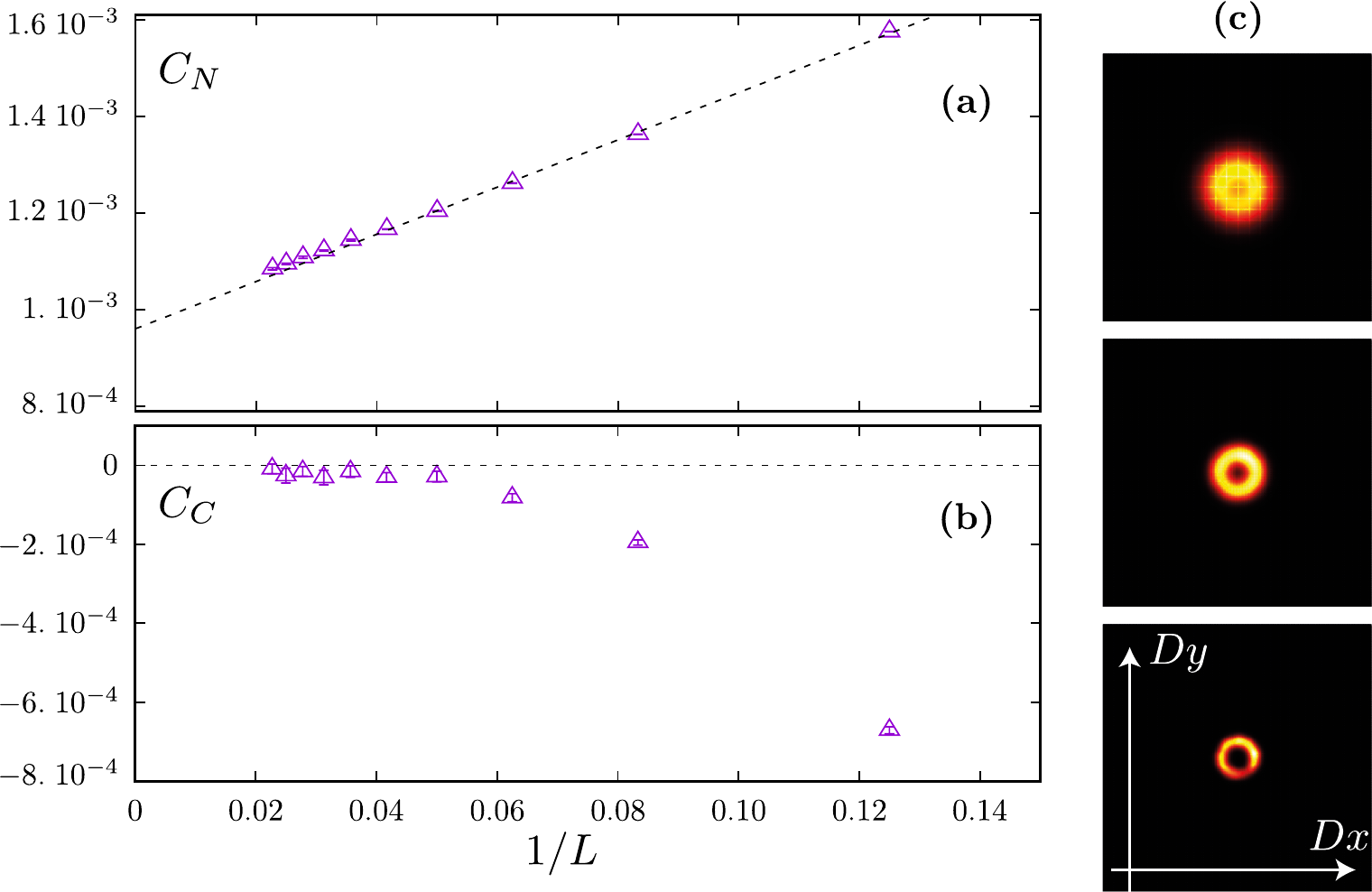}
\caption{
(a) Nematic correlator $C_N$ at separation $(L/2,L/2)$, extrapolated to a
non-zero value of $9.61(1)\times10^{-4}$ in the
thermodynamic limit for $\theta=-0.65\pi$ as a function of inverse size $1/L$.
(b) Same for the cartan correlation at separation $(L/2,L/2)$, see to vanish
for $1/L\to 0$.
(c) VBC histograms for sizes 16, 32 and 64 at $\theta=-0.5\pi$.\\
}
\label{fphases}
\end{figure}

{\it Long-range ordered phases -- }
To motivate the presence of nematic ordering in the range
$\theta\in[-0.75\pi,\theta_c]$, we present
Cartan and nematic correlation functions (defined below) for a system of linear
size $L$. We use three Cartan
 operators $C_{\alpha=1,2,3}=\sum_c b^c_\alpha | c \rangle \langle c | $ with  $b_1=\frac{1}{2}(1, 1, 0, 0, -1, -1), b_2=\frac{1}{2}(-1, 0, 1, -1, 0, 1),b_3=\frac{1}{2}(1, -1, 0, 0, 1, -1)$ corresponding to the underlying SU$(4)$ symmetry and diagonal in the color basis, forming the vector ${\bf C}=(C_1,C_2,C_3)$ at any site. To identify simple (anti-)ferromagnetic ordering, we consider the
Cartan correlator $C_C=\braket{ {\bf C}_{{\bf r}=(0,0)} \cdot {\bf C}_{{\bf r}=(L/2,L/2)} }$,
whereas  $C_N=\braket{Q_{1, {\bf r}=(0,0)} Q_{1,{\bf r}=(L/2,L/2)}}$, with the traceless operator $Q_1=C_1C_1-\frac{1}{6}$,
is used to identify nematic ordering. Details about the choice of Cartan
operators and connections to the spin operators of SU$(4)$ are provided in Sup. Mat.~\onlinecite{supmat}.

 Large size behaviors of these correlators are displayed in Fig.~\ref{fphases}(a,b) for $\theta=-0.65\pi$ (located in the nematic phase and relatively away from the critical point), where we clearly see that there
is long-range ordering in the nematic correlator but not in the Cartan correlator. We now turn to the VBC phase, which we first illustrate by the real space pattern (Fig.~\ref{fig:pd}) of dimer correlations, defined as $D_{\rm corr}(b)=
\braket{({\bf C}_{0,0}\cdot {\bf C}_{1,0})({\bf C}_{\vec{r}^b_1}\cdot {\bf C}_{\vec{r}^b_2})}$. Here $b$ indicates a bond number connecting nearest neighbor sites $\vec{r}^{\;b}_1$ and $\vec{r}^{\;b}_2$.
Data in Fig.~\ref{fig:pd} are taken at the SU($6$) point $\theta=-0.5\pi$ where previous simulations~\cite{harada2003neel} showed the existence of long-range VBC order, but without specification of the type of crystal encountered.
Note that we only present the connected correlation function, i.e. the value
$D_{\rm corr}^{\rm avg}=\frac{1}{N_b}\sum_b D_{\rm corr}(b)$ is subtracted out to only show
the non-trivial features.
An analysis of  the pattern in Fig.~\ref{fig:pd} along the lines of Ref.~\onlinecite{mambrini2006plaquette} reveals that it is different from the one expected in a pristine columnar state, but potentially  compatible with plaquette order. We provide next a detailed analysis of the symmetry of the VBC ordering.

{\it Emergence of a $U(1)$ symmetry -- }
For this, we define a vector order parameter
${\bf D}=(D_x,D_y)$ with $D_x=\sum_i (-1)^{i_x} C_{i_x,i_y}\cdot C_{i_x+1,i_y}$ and $D_y=\sum_i (-1)^{i_y} C_{i_x,i_y}\cdot C_{i_x,i_y+1}$.
We can build a 2D-histogram of ${\bf D}$ using the spatial configurations
generated in the QMC sampling. This is shown for the same parameter
values as in Fig.~\ref{fphases} where we clearly see a U$(1)$ symmetry emerging. A similar U$(1)$ symmetry is often observed for VBC phases close to DQCP~\cite{sandvik2007evidence,VBS-Sandvik,nahum_deconfined,pujari2013neel} but is generically not expected at the coexistence point between phases at a first order transition (see however recent works~\cite{zhao_symmetry-enhanced_2019,serna_O4,Takahashi}).
We find a finite order parameter for VBC order (characterized by a finite radius in Fig.~\ref{fphases}) and a U$(1)$ symmetry  (circular shape in  Fig.~\ref{fphases}) in the entire range  $[\theta_c,-\pi/2 ]$ on the system sizes accessible to us. We expect that eventually on larger sizes the histograms would show peaks at specific angles characteristic of the type of crystal ordering (e.g. at $0,\pm \pi/2,\pi$ for columnar order), but we are unable to reach this behavior. In the Sup.~Mat.\cite{supmat}, we present an analysis of the persistence of this U$(1)$ behavior for large $L$. We also expect the VBC to subsist for $\theta>-\pi/2$, even though it is difficult to pinpoint where it vanishes as QMC is not longer available.

{\it Weak first-order transition -- }
We now present evidence for a weak first-order transition between the nematic and VBC phases. Its weak nature makes it difficult to probe numerically, as several standard indications of a continuous phase transition are observed on small to intermediate length scales, as we now show.  As the nematic phase breaks a continuous SO$(6)$ symmetry, it is illuminating to carry out simulations in a basis where the symmetry is made explicit.
We call this basis the nematic basis (denoted by $\mN$) which is related to the sign-free color basis as follows:
$\ket{c}=\frac{1}{\sqrt{2}}(\ket{\mathbb{N}_c}-i\ket{\mathbb{N}_{\bar{c}}}),
\ket{\bar{c}}=\frac{1}{\sqrt{2}}(\ket{\mathbb{N}_c}+i\ket{\mathbb{N}_{\bar{c}}})$.
The Hamiltonian in this basis and the explicit $SO(6)$ symmetry are
detailed in Sup. Mat.~\cite{supmat}.
We can then define a 6-dimensional nematic order parameter
$M^c=\frac{1}{N}(\sum_i\ket{\mN_c}\bra{\mN_c})-\frac{1}{6}$, corresponding to "ferromagnetic" ordering in this basis.
The VBC ordering
is quantified by the amplitude of the VBC order parameter ${\bf D}^2=D_x^2+D_y^2$.

Given these order parameters, a traditional way of inquiring about the order of the phase transition is to consider their Binder cumulants. We find (see Sup. Mat.~\cite{supmat}) that while they clearly indicate the existence of long-range order away from the critical point, Binder cumulants have a non-monotonic behavior near $\theta_c$ which prevents for a conclusive determination of the nature of the phase transition.  We further consider the nematic ``color" stiffness $\rho_c$ defined using the spatial winding of loops in the QMC simulation as $\rho_c=\frac{1}{2}\braket{\sum_{\alpha=x,y} \sum_i(W^\alpha_i)^2}/\beta$, where $i$ runs over all the loops in a particular
space-time configuration. The spatial winding $W^\alpha_i$ of a particular loop $i$ is an integer counting how many sites it wraps over the
periodic boundary conditions of the system in the direction $\alpha$. This definition follows from a similar treatment of an SO$(3)$ system~\cite{kaul2012spin}.
We expect this stiffness to be finite in the nematic phase, to vanish in the VBC phase and to scale as $L^{-z}$ (with $z$ the dynamical critical exponent) at a continuous phase transition. Fig.~\ref{fPT} reveals a crossing of curves for different system sizes when rescaling the stiffness by $L$, which would be a signature of a continuous phase transition with $z=1$ close to $\theta\simeq -0.5969(1)\pi$.
This behavior is seen up to length scales of $L=36$. Further evidence for
behavior consistent with a continuous transition is provided by studies of the
second derivative of the local energy in Sup. Mat.~\cite{supmat} up to $L=36$ along with an estimate for the correlation length (effective) critical exponent $\nu$.
Detailed histograms for $L=32$ for the energy, nematic and VBC order parameters are also presented in Sup. Mat.~\cite{supmat} showing no discernible signatures of coexistence and hence compatible with a continuous transition up to this length scale.

\begin{figure}[t]
\includegraphics[width=\hsize]{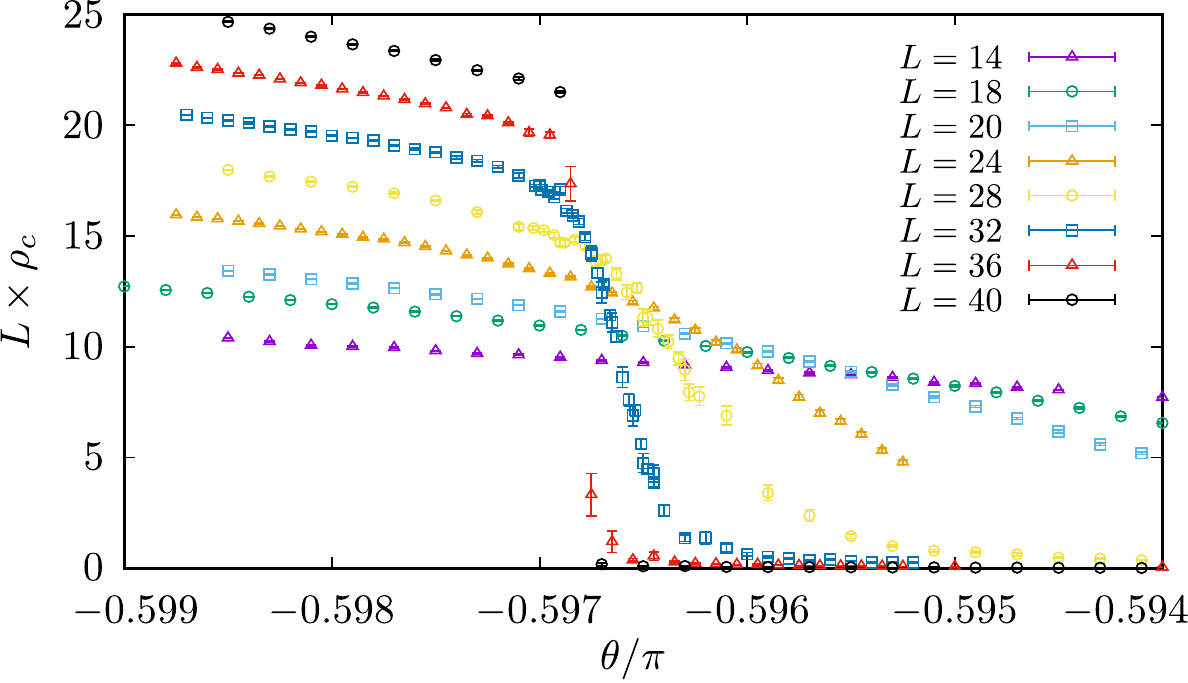}
\caption{
Nematic stiffness scaling across the transition. Here $\beta=2L$.}
\label{fPT}
\end{figure}

However, for larger sizes, we find a clear coexistence of both phases at the transition. This is shown in Fig.~\ref{ftrace} through a Monte Carlo time trace of the QMC data for a system with $L=40$. It can be seen that the system transits abruptly between the two phases,
consistent with the expectation for a first order transition. We have also
simulated system sizes up to $L=72$ and find that the jumps between
phases become increasingly unlikely with increasing size. Note that the
largest value that $(D_x^2+D_y^2)$ can take is $1$ for perfect
VBC ordering, compared to the value of $\approx0.007$ taken at the transition.
This indicates that the transition is only weakly first order and that it
cannot be identified for smaller sizes.
Note that as the nematic phase breaks a continuous symmetry, the values for
${\bf M}^2$ show a spread in Fig.~\ref{ftrace} but also in the nematic
phase. In the Sup. Mat.~\cite{supmat}, we also provide a comparison with the same transition occurring for the model Eq. \ref{eqHcolor} with $5$ colors, corresponding to an $SO(5)$ symmetry.

\begin{figure}[t]
\includegraphics[width=\hsize]{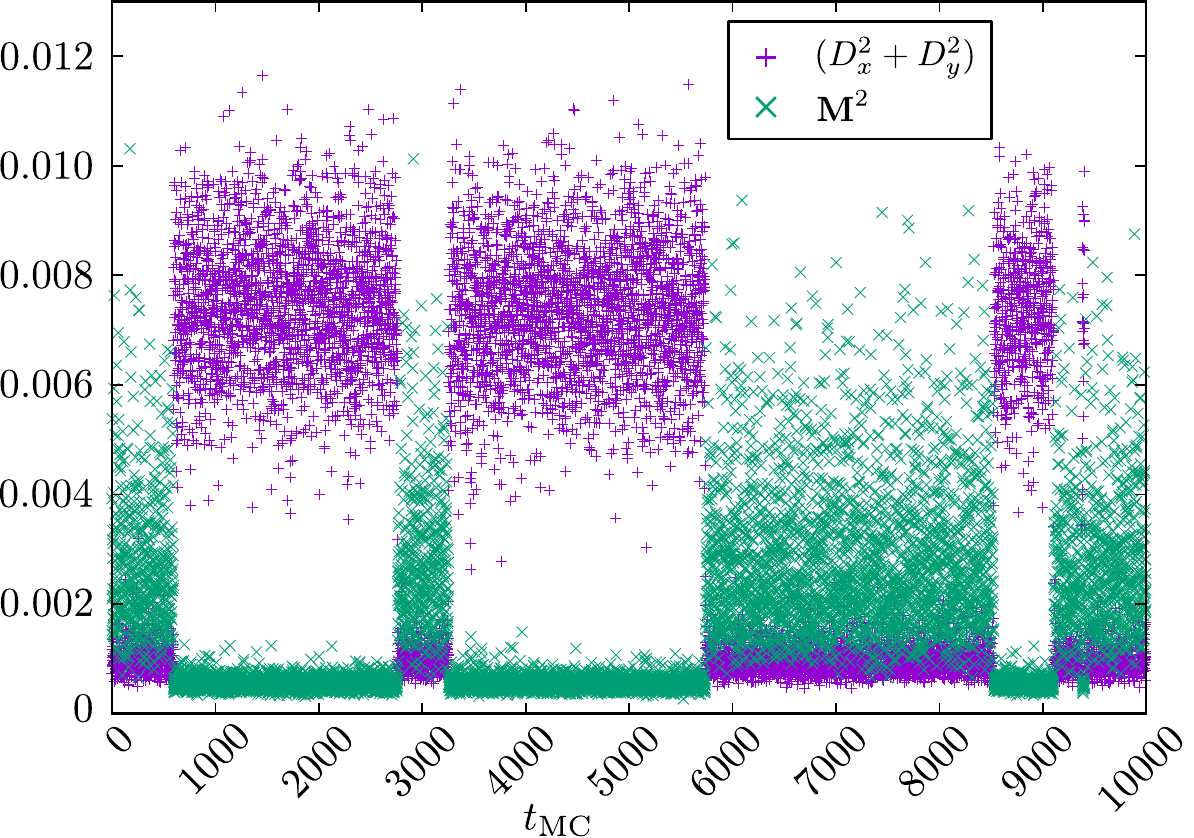}
\caption{Monte Carlo time series of the (square of the) order parameters at the best estimate
for the phase transition ($\theta=-0.5969\pi$) for $L=40$ and $\beta/L=0.4$.}
\label{ftrace}
\end{figure}

{\it Conclusion and perspectives -- }
In conclusion, using large-scale unbiased QMC simulations, we have shown the existence of a spin nematic phase bordered by a VBC phase (for $\theta > \theta_c$) and a ferromagnetic phase $(\theta < -3/4\pi)$ in a system of SU$(4)$ fermions with two particles per site. While the ferromagnetic/nematic transition is strongly first order (level crossings can be observed in exact diagonalization of small clusters~\cite{gauthe2020quantum}), we showed that the transition between  nematic and VBC phases is \emph{weakly first-order}. The relevance of biquadratic terms in cold-atomic systems~\cite{Yip2003,Imambekov2003,Eckert_2007,Brennen_2007,puetter2008theory} suggests that this model and its corresponding quantum phase transition can be realized in ultracold atomic setups. Note that a spin nematic phase has been observed in spin-1 spinor condensates~\cite{Zibold2016}.
The field theory analysis of Ref.~\cite{grover2007quantum}, written for SO($3$) spin-1 models on rectangular lattice, specifies that a continuous nematic-VBC transition is possible if double-instanton events are irrelevant at the transition point. The fact that our model is defined on a square lattice (where only four-fold instantons are allowed) and enjoys a higher SO($6$) symmetry (suggesting a higher scaling dimension of instantons events) hints at an even more likely occurence of a DQCP described by a similar field theory. We note that Ref.~\cite{grover2007quantum} predicts a U$(1)$ symmetry in the VBC order parameter, which we do observe in our simulations.
There are several reasons for a flow away from a putative DQCP. As mentioned in Ref.~\cite{grover2007quantum}, the U$(1)$ symmetry breaking operator can be relevant, which would cause a deviation from the DQCP. In our case, we do not see any evidence for a broken U$(1)$ at the length scales we can access. Another possibility would be that instabilities not present in the SO$(3)$ theory of the nematic to VBC transition for spin-$1$ systems are to be considered for the extended  SO$(6)$ symmetry present in the Hamiltonian studied in this work, calling for such a field theoretical analysis. Based on the above considerations, further fine-tuning of the weak first-order transition to a potential DQCP may be achieved by using another lattice (e.g. honeycomb), or by including diagonal bonds  (promoting plaquette order), or four-spin terms (favoring columnar order). While we were able to pinpoint the first-order nature of the transition in our work, in this perspective it would be useful to consider improved methods to probe weak first-order phase transitions, such as the recent proposal of Ref.~\onlinecite{DEmidio21}.  It is also interesting to contrast our results with those of recent studies~\cite{zhao_symmetry-enhanced_2019,serna_O4,Takahashi} observing emerging symmetries at weak first-order transitions in other models: we have checked that we do not find an enhanced symmetry between the VBC and nematic order parameters at $\theta_c$ (at least on the accessible lattice sizes).  Finally, we mention that the QMC algorithm in Sup. Mat.~\cite{supmat} (see, also, references \cite{sandvik1992generalization,albuquerque2010quantum,voll2015,keselman2020dimer,vollmayr_finite_1993} therein) allows to efficiently simulate bilinear-biquadratic SO($n_c$) models with arbitrary numbers of colors $n_c$, and for all lattices (including frustrated ones), with no sign problem in the range $\{ \theta_{\rm SF}\}$. Given the wide variety of exotic phases of matter including spin liquids that were encountered in previous studies of $SO(N)$ models with purely biquadratic interactions ($\theta=-\pi/2$)\cite{kaul2012spin,kaul2015spin,block_kagome_2020,wildeboer_first}, it thus paves the way for further fruitful explorations of exotic quantum physics in models with extended symmetries and competing energy scales.

\begin{acknowledgements}

We thank D. Poilblanc for useful discussions and collaboration on related work.  This work benefited from the support of the project LINK ANR-18-CE30-0022-04 of the French National Research Agency (ANR). We acknowledge the use of HPC resources from CALMIP (grants 2020-P0677 and 2021-P0677) and GENCI (grant x2021050225). We use the ALPS library~\cite{ALPS13,ALPS2} for some of our QMC simulations.

\end{acknowledgements}

\bibliography{SU4}

\pagebreak
\clearpage

\begin{widetext}

\begin{center}
\textbf{Supplemental Material for  ``Weakly first-order quantum phase transition between Spin Nematic and Valence Bond Crystal Order in a square lattice SU(4) fermionic model"}
\end{center}

\end{widetext}

\author{Pranay Patil}
\affiliation{Laboratoire de Physique Th\'eorique, Universit\'e de Toulouse, CNRS, UPS, France}

\author{Fabien Alet}
\affiliation{Laboratoire de Physique Th\'eorique, Universit\'e de Toulouse, CNRS, UPS, France}

\author{Sylvain Capponi}
\affiliation{Laboratoire de Physique Th\'eorique, Universit\'e de Toulouse, CNRS, UPS, France}

\author{Matthieu Mambrini}
\affiliation{Laboratoire de Physique Th\'eorique, Universit\'e de Toulouse, CNRS, UPS, France}

%

\maketitle

\section{Derivation of the sign-free SO$(6)$ color Hamiltonian from the SU$(4)$ fermionic Hamiltonian  }

\begin{table}
{\footnotesize  $
\begin{array}{l}
        S^{(1)}=\mathbb{X}_1=\frac{1}{\sqrt{2}} \left(
        \begin{array}{cccccc}
                0 & 0 & 0 & 0 & 0 & 0 \\
                0 & 0 & 1 & 0 & 0 & 0 \\
                0 & 1 & 0 & 0 & 0 & 0 \\
                0 & 0 & 0 & 0 & 1 & 0 \\
                0 & 0 & 0 & 1 & 0 & 0 \\
                0 & 0 & 0 & 0 & 0 & 0 \\
        \end{array}
        \right) \\
        S^{(2)}=\mathbb{X}_2=\frac{1}{\sqrt{2}} \left(
        \begin{array}{cccccc}
                0 & 0 & 1 & 0 & 0 & 0 \\
                0 & 0 & 0 & 0 & 0 & 0 \\
                1 & 0 & 0 & 0 & 0 & 0 \\
                0 & 0 & 0 & 0 & 0 & -1 \\
                0 & 0 & 0 & 0 & 0 & 0 \\
                0 & 0 & 0 & -1 & 0 & 0 \\
        \end{array}
        \right) \\
        S^{(3)}=\mathbb{X}_3=\frac{1}{\sqrt{2}} \left(
        \begin{array}{cccccc}
                0 & 0 & 0 & 0 & -1 & 0 \\
                0 & 0 & 0 & 0 & 0 & -1 \\
                0 & 0 & 0 & 0 & 0 & 0 \\
                0 & 0 & 0 & 0 & 0 & 0 \\
                -1 & 0 & 0 & 0 & 0 & 0 \\
                0 & -1 & 0 & 0 & 0 & 0 \\
        \end{array}
        \right) \\
        S^{(4)}=\mathbb{X}_4=\frac{1}{\sqrt{2}} \left(
        \begin{array}{cccccc}
                0 & 1 & 0 & 0 & 0 & 0 \\
                1 & 0 & 0 & 0 & 0 & 0 \\
                0 & 0 & 0 & 0 & 0 & 0 \\
                0 & 0 & 0 & 0 & 0 & 0 \\
                0 & 0 & 0 & 0 & 0 & 1 \\
                0 & 0 & 0 & 0 & 1 & 0 \\
        \end{array}
        \right) \\
        S^{(5)}=\mathbb{X}_5=\frac{1}{\sqrt{2}} \left(
        \begin{array}{cccccc}
                0 & 0 & 0 & -1 & 0 & 0 \\
                0 & 0 & 0 & 0 & 0 & 0 \\
                0 & 0 & 0 & 0 & 0 & 1 \\
                -1 & 0 & 0 & 0 & 0 & 0 \\
                0 & 0 & 0 & 0 & 0 & 0 \\
                0 & 0 & 1 & 0 & 0 & 0 \\
        \end{array}
        \right) \\
        S^{(6)}=\mathbb{X}_6=\frac{1}{\sqrt{2}} \left(
        \begin{array}{cccccc}
                0 & 0 & 0 & 0 & 0 & 0 \\
                0 & 0 & 0 & 1 & 0 & 0 \\
                0 & 0 & 0 & 0 & 1 & 0 \\
                0 & 1 & 0 & 0 & 0 & 0 \\
                0 & 0 & 1 & 0 & 0 & 0 \\
                0 & 0 & 0 & 0 & 0 & 0 \\
        \end{array}
        \right) \\
        S^{(7)}=\mathbb{Y}_1=\frac{1}{\sqrt{2}} \left(
        \begin{array}{cccccc}
                0 & 0 & 0 & 0 & 0 & 0 \\
                0 & 0 & -i & 0 & 0 & 0 \\
                0 & i & 0 & 0 & 0 & 0 \\
                0 & 0 & 0 & 0 & -i & 0 \\
                0 & 0 & 0 & i & 0 & 0 \\
                0 & 0 & 0 & 0 & 0 & 0 \\
        \end{array}
        \right) \\
        S^{(8)}=\mathbb{Y}_2=\frac{1}{\sqrt{2}} \left(
        \begin{array}{cccccc}
                0 & 0 & -i & 0 & 0 & 0 \\
                0 & 0 & 0 & 0 & 0 & 0 \\
                i & 0 & 0 & 0 & 0 & 0 \\
                0 & 0 & 0 & 0 & 0 & i \\
                0 & 0 & 0 & 0 & 0 & 0 \\
                0 & 0 & 0 & -i & 0 & 0 \\
        \end{array}
        \right) \\
\end{array}
        $}
\caption{SU(4) generators of the $\protect\su{0}{1,1}{6}$ representation in basis ${\cal B}$.}
\label{tab:generators1}
\end{table}

\begin{table}
{\footnotesize  $
        \begin{array}{l}
                 S^{(9)}=\mathbb{Y}_3=\frac{1}{\sqrt{2}} \left(
                \begin{array}{cccccc}
                        0 & 0 & 0 & 0 & i & 0 \\
                        0 & 0 & 0 & 0 & 0 & i \\
                        0 & 0 & 0 & 0 & 0 & 0 \\
                        0 & 0 & 0 & 0 & 0 & 0 \\
                        -i & 0 & 0 & 0 & 0 & 0 \\
                        0 & -i & 0 & 0 & 0 & 0 \\
                \end{array}
                \right) \\
                S^{(10)}=\mathbb{Y}_4=\frac{1}{\sqrt{2}} \left(
                \begin{array}{cccccc}
                        0 & -i & 0 & 0 & 0 & 0 \\
                        i & 0 & 0 & 0 & 0 & 0 \\
                        0 & 0 & 0 & 0 & 0 & 0 \\
                        0 & 0 & 0 & 0 & 0 & 0 \\
                        0 & 0 & 0 & 0 & 0 & -i \\
                        0 & 0 & 0 & 0 & i & 0 \\
                \end{array}
                \right) \\
                S^{(11)}=\mathbb{Y}_5=\frac{1}{\sqrt{2}} \left(
                \begin{array}{cccccc}
                        0 & 0 & 0 & i & 0 & 0 \\
                        0 & 0 & 0 & 0 & 0 & 0 \\
                        0 & 0 & 0 & 0 & 0 & -i \\
                        -i & 0 & 0 & 0 & 0 & 0 \\
                        0 & 0 & 0 & 0 & 0 & 0 \\
                        0 & 0 & i & 0 & 0 & 0 \\
                \end{array}
                \right) \\
                S^{(12)}=\mathbb{Y}_6=\frac{1}{\sqrt{2}} \left(
                \begin{array}{cccccc}
                        0 & 0 & 0 & 0 & 0 & 0 \\
                        0 & 0 & 0 & -i & 0 & 0 \\
                        0 & 0 & 0 & 0 & -i & 0 \\
                        0 & i & 0 & 0 & 0 & 0 \\
                        0 & 0 & i & 0 & 0 & 0 \\
                        0 & 0 & 0 & 0 & 0 & 0 \\
                \end{array}
                \right) \\
                S^{(13)}= \mathbb{Z}_1=\left(
                \begin{array}{cccccc}
                        0 & 0 & 0 & 0 & 0 & 0 \\
                        0 & \frac{1}{\sqrt{2}} & 0 & 0 & 0 & 0 \\
                        0 & 0 & \frac{1}{\sqrt{2}} & 0 & 0 & 0 \\
                        0 & 0 & 0 & -\frac{1}{\sqrt{2}} & 0 & 0 \\
                        0 & 0 & 0 & 0 & -\frac{1}{\sqrt{2}} & 0 \\
                        0 & 0 & 0 & 0 & 0 & 0 \\
                \end{array}
                \right) \\
                S^{(14)}= \mathbb{Z}_2=\left(
                \begin{array}{cccccc}
                        \sqrt{\frac{2}{3}} & 0 & 0 & 0 & 0 & 0 \\
                        0 & -\frac{1}{\sqrt{6}} & 0 & 0 & 0 & 0 \\
                        0 & 0 & \frac{1}{\sqrt{6}} & 0 & 0 & 0 \\
                        0 & 0 & 0 & -\frac{1}{\sqrt{6}} & 0 & 0 \\
                        0 & 0 & 0 & 0 & \frac{1}{\sqrt{6}} & 0 \\
                        0 & 0 & 0 & 0 & 0 & -\sqrt{\frac{2}{3}} \\
                \end{array}
                \right) \\
                S^{(15)}= \mathbb{Z}_3=\left(
                \begin{array}{cccccc}
                        \frac{1}{\sqrt{3}} & 0 & 0 & 0 & 0 & 0 \\
                        0 & \frac{1}{\sqrt{3}} & 0 & 0 & 0 & 0 \\
                        0 & 0 & -\frac{1}{\sqrt{3}} & 0 & 0 & 0 \\
                        0 & 0 & 0 & \frac{1}{\sqrt{3}} & 0 & 0 \\
                        0 & 0 & 0 & 0 & -\frac{1}{\sqrt{3}} & 0 \\
                        0 & 0 & 0 & 0 & 0 & -\frac{1}{\sqrt{3}} \\
                 \end{array}
                 \right) \\
                 \end{array}
                 $}

\caption{SU(4) generators of the $\protect\su{0}{1,1}{6}$ representation in basis ${\cal B}$ (continued from Tab.\ref{tab:generators1}).}
\label{tab:generators2}
\end{table}

The {\bf 6}-representation of SU(4), corresponding to the $\su{0}{1,1}{6}$
Young tableau, can be interpreted as the onsite Hilbert space of a pair of
fermions or a 6-component SU(4) spin. We refer to the basis
${\cal B}=\{\vert 1 \rangle,\ldots,\vert 6 \rangle\}$ as the original basis in
the following. The bilinear-biquadratic model studied in this work is
defined using the spin operator ${\bf S}=\{S^{(\alpha)}\}$ which is a 15-
component vector formed by the generators in the considered representation of
SU$(4)$. In analogy with SU$(2)$ -- where the generators are $S^x$
(real symmetric), $S^y$ (imaginary antisymmetric) and $S^z$ (diagonal) --
we use the alternative notation  $\mathbb{X}_1,\ldots,\mathbb{X}_6$ for
$S^{(1)},\ldots,S^{(6)}$, $\mathbb{Y}_1,\ldots,\mathbb{Y}_6$ for
$S^{(7)},\ldots,S^{(12)}$ and $\mathbb{Z}_1,\ldots,\mathbb{Z}_3$ for
$S^{(13)},\ldots,S^{(15)}$. The convention used in this paper for the matrix
representation of these generators  is given in Tables~\ref{tab:generators1}
and \ref{tab:generators2}.

The diagonal and off-diagonal matrix elements of the two-site Hamiltonian ${\cal H} -(K/4)      \mathds{1}$ in the basis ${\cal B}$ are respectively given by
\begin{equation*}
        {\footnotesize
        \begin{array}[t]{ccc}
                {\text{\bf Diagonal}} && {\text{\bf Off-diagonal}}\\
\begin{array}[t]{c|c}\hline\hline
        J & K-J \\ \hline
        \begin{array}{c}
                s_1\otimes s_1 \\
                s_2\otimes s_2 \\
                s_3\otimes s_3 \\
                s_4\otimes s_4 \\
                s_5\otimes s_5 \\
                s_6\otimes s_6 \\
        \end{array}
        &
        \begin{array}{c}
                s_1\otimes s_6 \\
                s_2\otimes s_5 \\
                s_3\otimes s_4 \\
                s_4\otimes s_3 \\
                s_5\otimes s_2 \\
                s_6\otimes s_1 \\
        \end{array}
        \\ \hline\hline
\end{array} &&
                \begin{array}[t]{c|c}\hline\hline
                        J & K   \\ \hline
                        \begin{array}[t]{cc}
                                s_1\otimes s_2 & s_2\otimes s_1 \\
                                s_1\otimes s_3 & s_3\otimes s_1 \\
                                s_1\otimes s_4 & s_4\otimes s_1 \\
                                s_1\otimes s_5 & s_5\otimes s_1 \\
                                s_2\otimes s_3 & s_3\otimes s_2 \\
                                s_2\otimes s_4 & s_4\otimes s_2 \\
                                s_2\otimes s_6 & s_6\otimes s_2 \\
                                s_3\otimes s_5 & s_5\otimes s_3 \\
                                s_3\otimes s_6 & s_6\otimes s_3 \\
                                s_4\otimes s_5 & s_5\otimes s_4 \\
                                s_4\otimes s_6 & s_6\otimes s_4 \\
                                s_5\otimes s_6 & s_6\otimes s_5 \\
                        \end{array}
                        &
                        \begin{array}[t]{cc}
                                s_1\otimes s_6 & s_6\otimes s_1 \\
                                s_2\otimes s_5 & s_5\otimes s_2 \\
                                s_3\otimes s_4 & s_4\otimes s_3 \\
                        \end{array} \\ \hline\hline
                        J-K & K-J \\ \hline
                        \begin{array}[t]{cc}
                                s_1\otimes s_6 & s_2\otimes s_5 \\
                                s_1\otimes s_6 & s_5\otimes s_2 \\
                                s_2\otimes s_5 & s_3\otimes s_4 \\
                                s_2\otimes s_5 & s_4\otimes s_3 \\
                                s_2\otimes s_5 & s_6\otimes s_1 \\
                                s_3\otimes s_4 & s_5\otimes s_2 \\
                                s_4\otimes s_3 & s_5\otimes s_2 \\
                                s_5\otimes s_2 & s_6\otimes s_1 \\
                        \end{array}
                        &
                        \begin{array}[t]{cc}
                                s_1\otimes s_6 & s_3\otimes s_4 \\
                                s_1\otimes s_6 & s_4\otimes s_3 \\
                                s_3\otimes s_4 & s_6\otimes s_1 \\
                                s_4\otimes s_3 & s_6\otimes s_1 \\
                        \end{array}
                        \\ \hline\hline
                \end{array}
\end{array}
}
\end{equation*}
where we use the ``state" notation $s_i$ for $| i \rangle$.

In this basis ${\cal B}$, as seen in the above table, the Hamiltonian suffers from a sign problem except when $J=K\leq0$ which corresponds to one SU(6) point. From the above table one immediately notice that, for $J=K$, the Hamiltonian is just a six-color exchange model on the two interacting sites.

Interestingly, the range of parameters for which the model is sign-free can be extended to a finite range of $\{J,K\}$ including the two SU(6)-symmetric points $\bf{6}-\bf{6}$ ($J=K<0$) and $\bf{6}-\bf{\bar{6}}$ ($J=0,K<0$).

The sign-free basis ${\cal C}$ (for ``color" basis) is constructed by a simple redefinition of the six states of ${\cal B}$ :

\begin{align}
        \label{eq:signfreebasis}
        s_1 &= -c_3 \\
        s_2 &= -c_1 \nonumber \\
        s_3 &= c_2  \nonumber \\
        s_4 &= c_5  \nonumber \\
         s_5 &= c_6 \nonumber \\
         s_6 &= -c_4 \nonumber
\end{align}

Basically, this transformation merges the $J-K$ and $K-J$ off-diagonal amplitudes of the model in the original basis $\cal B$, as can be seen from the
matrix elements in the new basis $\cal C$ :

\begin{equation*}
        {\footnotesize
                \begin{array}[t]{ccc}
                        {\text{\bf Diagonal}} && {\text{\bf Off-diagonal}}\\
                        \begin{array}[t]{c|c}\hline\hline
                                J & K-J \\ \hline
                                \begin{array}{c}
                                        c_1\otimes c_1 \\
                                        c_2\otimes c_2 \\
                                        c_3\otimes c_3 \\
                                        c_4\otimes c_4 \\
                                        c_5\otimes c_5 \\
                                        c_6\otimes c_6 \\
                                \end{array}
                                &
                        \begin{array}{c}
                                c_1\otimes c_6 \\
                                c_2\otimes c_5 \\
                                c_3\otimes c_4 \\
                                c_4\otimes c_3 \\
                                c_5\otimes c_2 \\
                                c_6\otimes c_1 \\
                        \end{array}
                                \\ \hline\hline
                        \end{array} &&
                        \begin{array}[t]{c|c}\hline\hline
                                J & K-J   \\ \hline
                \begin{array}[t]{cc}
                        c_1\otimes c_2 & c_2\otimes c_1 \\
                        c_1\otimes c_3 & c_3\otimes c_1 \\
                        c_1\otimes c_4 & c_4\otimes c_1 \\
                        c_1\otimes c_5 & c_5\otimes c_1 \\
                        c_2\otimes c_3 & c_3\otimes c_2 \\
                        c_2\otimes c_4 & c_4\otimes c_2 \\
                        c_2\otimes c_6 & c_6\otimes c_2 \\
                        c_3\otimes c_5 & c_5\otimes c_3 \\
                        c_3\otimes c_6 & c_6\otimes c_3 \\
                        c_4\otimes c_5 & c_5\otimes c_4 \\
                        c_4\otimes c_6 & c_6\otimes c_4 \\
                        c_5\otimes c_6 & c_6\otimes c_5 \\
                       {\color{blue} (c_1\otimes c_6} &        {\color{blue} c_6\otimes c_1)}\\
                        {\color{blue} (c_2\otimes c_5} &        {\color{blue} c_5\otimes c_2) }\\
                        {\color{blue} (c_3\otimes c_4} &        {\color{blue} c_4\otimes c_3)} \\
                \end{array}
                                &
                        \begin{array}[t]{cc}
                                c_1\otimes c_6 & c_2\otimes c_5 \\
                                c_1\otimes c_6 & c_3\otimes c_4 \\
                                c_1\otimes c_6 & c_4\otimes c_3 \\
                                c_1\otimes c_6 & c_5\otimes c_2 \\
                                c_2\otimes c_5 & c_3\otimes c_4 \\
                                c_2\otimes c_5 & c_4\otimes c_3 \\
                                c_2\otimes c_5 & c_6\otimes c_1 \\
                                c_3\otimes c_4 & c_5\otimes c_2 \\
                                c_3\otimes c_4 & c_6\otimes c_1 \\
                                c_4\otimes c_3 & c_5\otimes c_2 \\
                                c_4\otimes c_3 & c_6\otimes c_1 \\
                                c_5\otimes c_2 & c_6\otimes c_1 \\
                        \end{array} \\ \hline\hline
                        \multicolumn{2}{c}{K {\color{blue} (-J)}} \\ \hline
                                \multicolumn{2}{c}{\begin{array}{cc}
                                                c_1\otimes c_6 & c_6\otimes c_1 \\
                                                c_2\otimes c_5 & c_5\otimes c_2 \\
                                                c_3\otimes c_4 & c_4\otimes c_3 \\
                                \end{array}}                            \\ \hline\hline
                        \end{array}
                \end{array}
        }
\end{equation*}
A simple inspection at the right column of the above table shows that the sign free condition in $\cal C$ is now $J<0$, $K<0$, and $K<J$.

Let us remark that the basis change (\ref{eq:signfreebasis}) is a uniform on-site transformation that does not require any  hypothesis  about the bipartite nature of the lattice.

Adding the blue parenthesized (summing to zero) terms in the above table, and introducing the notation $\bar{c}_i=c_{7-i}$ leads to a more compact (and suitable for the quantum Monte-Carlo algorithm presented later) expression for the Hamiltonian in basis $\cal C$:

\begin{equation}
        {\cal H} = \sum_{\langle i,j\rangle} \left( J \sum_{c,c'} | cc' \rangle \langle c' c| + (K-J) \sum_{c,s} | c\bar{c} \rangle \langle s \bar{s}| \right) + \frac{K}{4} \mathds{1},
        \label{eqHcolor}
\end{equation}
which is the form presented in the main text, up to the irrelevant constant $K/4$.
In the case of SU(2) spins, (anti)ferromagnetic order can be probed using two-point $S_z$ correlations $\langle S^z_i S^z_j \rangle$. The SU(4) generalization involves the 3 generators of the Cartan subalgebra. Among the possible choices, we can consider the natural set $\{\mathbb{Z}_1, \mathbb{Z}_2,\mathbb{Z}_3\}$ or the one adopted in the main text $\{C_1,C_2,C_3 \}$. Of course these two sets carry the same information and are simply related by linear relations:
\begin{align}
        \label{eq:ZtoC}
        C_1 &=   \sqrt{2}  \mathbb{Z}_1 \\ \nonumber
        C_2 &=  -\frac{1}{2 \sqrt{2}}  \mathbb{Z}_1  +\frac{1}{2} \sqrt{\frac{3}{2}} \mathbb{Z}_2\\
        C_3 &=   -\frac{1}{\sqrt{6}} \mathbb{Z}_2 +\frac{1}{\sqrt{3}}  \nonumber \mathbb{Z}_3.
\end{align}

\section{Quantum Monte Carlo loop algorithm for the bilinear-biquadratic SO($6$) 6-color Hamiltonian  }

This section details how to implement an efficient cluster quantum Monte Carlo algorithm for the SO($6$) Hamiltonian \eqref{eqHcolor}. The algorithm presented below is a simple adaption of the so-called non-binary loop algorithm proposed by Kawashima and Harada~\cite{kawashima2004} for bilinear-biquadratic spin 1 models in the region $\theta \in [-3/4\pi,-\pi/2]$. We present it using the Stochastic Series Expansion~\cite{sandvik1992generalization} framework, and considering an arbitrary number of colors $n_c$ in its construction, meaning that it can be applied directly for the same SO($n_c$) Hamiltonian (we specialized to $n_c=6$ in the simulations presented in the main text).

As Stochastic Series Expansion calculates expectation values by sampling over
operator strings generated upon expanding $\mathrm{Tr}[e^{-\beta H}]$, we seek
a convenient representation for the operator string.
We decompose the Hamiltonian given in Eq.~\eqref{eqHcolor} as
${\cal H}=-\sum_{b=\langle i,j\rangle} H_{b}^1 + H_{b}^2$ with $H_{b}^1=|J| \sum_{c,c'} | cc' \rangle \langle c' c|$ and $H_{b}^2=(J-K) \sum_{c,s} | c\bar{c} \rangle \langle s \bar{s}|$.
An operator and its matrix element $\langle c_1 c_2 | H_b^{1/2} | c'_1 c'_2 \rangle$ can be represented as a vertex with four legs $(c_1,c_2,c'_1,c'_2)$ as shown below. The two types of terms in the decomposition of the Hamiltonian encode different constraints on these legs, and can be represented as
a {\it cross} graph for the first term $H_{b}^1$ and
a {\it horizontal} graph for the second $H_{b}^2$:
\begin{center}
$
\begin{tikzpicture}[baseline={([xshift=6pt,yshift=0pt]current bounding box.center)},scale=0.4,] \draw [thick,dashed]  (-0.6,-0.6)   -- (-0.6,0.6); \draw [thick,dashed]  (-0.6,0.6)   -- (0.6,0.6);\draw [thick,dashed] (0.6,0.6)   -- (0.6,-0.6);\draw [thick,dashed]  (0.6,-0.6)   -- (-0.6,-0.6); \node at (-0.7,-1.1) {$c_1$};\node at (0.7,1.2) {$c_2^{\prime}$} ;\node at (-0.7,1.2) {$c_1^{\prime}$} ;\node at (0.7,-1.1) {$c_2$};  \end{tikzpicture}
\ \ \ \ \ \ \ \ \ \ \ \ \ \ \ \ \ H_b^1 \ \ \ \ \hat{=} \ \ \ \ \begin{tikzpicture}[baseline={([xshift=3pt,yshift=0pt]current bounding box.center)},scale=0.4,] \draw [thick]  (-0.5,-0.5)   -- (0.5,0.5); \draw [thick]  (-0.5,0.5)   -- (0.5,-0.5); \node at (-0.7,-0.81) {$c$} ;\node at (0.61,0.81) {$c$} ;\node at (-0.5,0.95) {$c^{\prime}$} ;\node at (0.7,-0.68) {$c^{\prime}$}; \end{tikzpicture}
\ \ \ \ \ \ \ \ \ H_b^2 \ \ \ \ \hat{=} \ \ \ \ \begin{tikzpicture}[baseline={([xshift=6pt,yshift=0pt]current bounding box.center)},scale=0.4,] \draw [thick]  (-0.5,-0.5)   -- (-0.5,-0.3); \draw [thick]  (-0.5,-0.3)   -- (0.5,-0.3);\draw [thick] (0.5,-0.3)   -- (0.5,-0.5);\draw [thick]  (-0.5,0.5)   -- (-0.5,0.3); \draw [thick]  (-0.5,0.3)   -- (0.5,0.3);\draw [thick] (0.5,0.3)   -- (0.5,0.5);\node at (-0.5,-0.93) {$c$};\node at (0.5,0.93) {$\bar{s}$} ;\node at (-0.5,0.9) {$s$} ;\node at (0.5,-0.9) {$\bar{c}$};  \end{tikzpicture}
$
\end{center}
where sums over indices are implied. In addition to these operators we add an
identity operator, $I_i$ indexed by site number $i$,
which allows us to implement efficient updating methods.
Using this notation, an operator string
such as $...H_{b_1}^1H_{b_2}^2H_{b_3}^2H_{b_4}^1H_{b_5}^2I_{_6}H_{b_7}^1...$
would
map to a configuration of vertices dictated by the rules discussed above. This
can also be seen as a loop configuration by connecting the legs
of vertices occuring sequentially in the operator string. This is a well
established procedure for quantum Monte Carlo and examples of such loop
configurations can be found in Ref.~\cite{kawashima2004}.
Starting
from a random operator string, we can sample relevant operator strings using
the two following steps of the algorithm:\\

\noindent
\textit{Diagonal update:}
The diagonal elements of the Hamiltonian can be inserted/removed in the diagonal update.
When an identity operator is encountered, one proposes to insert a diagonal operator on a random bond with a probability $\frac{2NK\beta}{M-m}$, where $N$ is
the number of lattice sites, $m$ is the current number of non-identity
operators, and $M$ is the fixed cutoff for the operator string length which
is set to be large enough to accomodate all fluctuations of $m$.
\cite{sandvik1992generalization}.

Only  the following two situations (for even number of colors $n_c$) for the colors of the currently propagated states lead to an insertion:
\begin{itemize}
\item 1. If the colors are identical $(c=c')$, one proposes to insert a $H^1_b$ opertor with a probability proportional to the matrix element $|J|$.
\item 2. If the colors are complementary $(c=\bar{c})$, one proposes to insert a $H^2_b$ operator using the matrix element $|K-J|$.
\end{itemize}
For an odd number of colors, one must be careful to consider the case of
$c=c'=\bar{c}$ separately, as both types of operators have non-zero matrix
elements in this case, and the probability of addition should be proportional
to $|K|$.

When a diagonal operator is encountered, it is removed with probability
$\frac{M-m+1}{2NK\beta}$.

\textit{Loop update:}
A loop is sourced by picking a leg of a vertex at random (which has a color $c_0$), and propagating a loop of randomly selected color $c_l \neq c_0$.
When the loop hits a vertex on a certain leg ({\it e.g.} leg with state $c_1$ as shown in the diagram above) of a vertex,
it will first change the color $c_1 \rightarrow c_l$ and continue its path using different moves depending on the type of vertex encountered:\\
\noindent
1. Cross vertices: When $c_1 = c'_2$ (but $c_2 \neq \bar{c_1}$), the loop does a {\it diagonal} move $c'_2 \rightarrow c_l$ and continues propagating (with color $c_l$)\\
\noindent
2. Horizontal vertices: When $c_1 = \bar{c_2}$ (but $c'_2 \neq c_1$), the loop reverses its direction and color $c_l \rightarrow \bar{c_l}$, switches $c_2   \rightarrow \bar{c_l}$ and continues propagating  (with color $\bar{c_l}$)\\
\noindent
3. Mixed vertices: When $c_1 = c'_2 = \bar{c_2}$, then with probability $p_{\rm diag}=J/K$ the loop does a {\it diagonal move} (move 1), and with probability $1-p_{\rm diag}$ switches and reverses (move 2).\\
The loop goes on until it reaches its initial starting point. This loop is accepted with probability one.\\

At $\theta=-\pi/2$ and for bipartite lattices, the model is SU($n_c$)-symmetric and the algorithm is identical to the one derived for SU($N$) models \cite{kaul2011quantum,beach2009n}.
For $\theta=-3/4 \pi$, the model is also SU($n_c$)-symmetric (with fundamental representation on each lattice site).
Quite importantly, the algorithm is not dependent on the bipartite nature of the lattice and can thus be applied to any arbitrary lattice. A special case of the algorithm at $\theta=-\pi/2$ has been used for studies of SO($3$) triangular lattice models, and SO($n$) models on kagome and triangular lattices~\cite{kaul2012spin,kaul2015spin,block_kagome_2020}. Note also Ref.~\cite{voll2015} which studies the spin-1 bilinear-biquadratic model on triangular lattice, using a similar 3-color loop algorithm in the region $\{\theta_{\rm SF}\}$.

\section{Mapping to Nematic Hamiltonian and SO$(6)$ symmetry}

To make the nematic ordering generated by Hamiltonian \eqref{eqHcolor} more
explicit, we first reproduce the transformation of the color states to the nematic basis from the main paper:

\be
\ket{c}=\frac{1}{\sqrt{2}}(\ket{\mathbb{N}_c}-i\ket{\mathbb{N}_{\bar{c}}}),\
\ket{\bar{c}}=\frac{1}{\sqrt{2}}(\ket{\mathbb{N}_c}+i\ket{\mathbb{N}_{\bar{c}}}),\
\ee
Using these relations and noting that the second term in Eq.~\eqref{eqHcolor}
can be written as
$\big(\sum_c\ket{c\bar{c}}\big)\big(\sum_s\bra{s\bar{s}}\big)$,
a simple substitution shows that
$\ket{c\bar{c}}+\ket{\bar{c}c}=\ket{\mN_c\mN_c}+\ket{\mN_{\bc}\mN_{\bc}}$,
leading to
\be
\big(\sum_c\ket{c\bar{c}}\big)\big(\sum_s\bra{s\bar{s}}\big)
=\big(\sum_c\ket{\mN_c\mN_c}\big)\big(\sum_s\bra{\mN_s\mN_s}\big).
\ee

To transform the first term in Eq.~\eqref{eqHcolor}, we first note that the
36 terms in the complete sum over $c,\cp$ can be separated into sets of 4,
each given by
\be\label{eq4set}
\ket{c\cp}\bra{\cp c}+
\ket{\bc\cp}\bra{\cp\bc}+
\ket{c\bcp}\bra{\bcp c}+
\ket{\bc\bcp}\bra{\bcp\bc}.
\ee
Doing the transformation on the first two terms and only on the first site,
we see that $\ket{c\cp}\bra{\cp c}+\ket{\bc\cp}\bra{\cp\bc}=
\ket{\mN_c\cp}\bra{\cp\mN_c}+\ket{\mN_{\bc}\cp}\bra{\cp\mN_{\bc}}.$
Following this with the same transformation for the last two terms, a subsequent
transformation of the second site, and a careful counting of remaining terms
leads to Eq.~\eqref{eq4set} being expressed in the nematic basis as
\begin{multline}
\label{eq4setN}
\ket{\mN_{c}\mN_{\cp}}\bra{\mN_{\cp} \mN_{c}}+
\ket{\mN_{\bc}\mN_{\cp}}\bra{\mN_{\cp} \mN_{\bc}}+
\ket{\mN_{c}\mN_{\bcp}}\bra{\mN_{\bcp} \mN_{c}}\\+
\ket{\mN_{\bc}\mN_{\bcp}}\bra{\mN_{\bcp} \mN_{\bc}}.
\end{multline}
As one can see from the above equation, this term retains the same form in the
nematic basis. The complete Hamiltonian in this basis is expressed as
\begin{multline}\label{eqHnem}
H=\sum_{\langle i,j\rangle} J \sum_{c,c'} |\mathbb{N}_c\mathbb{N}_{c'} \rangle \langle \mathbb{N}_{c'}\mathbb{N}_c| \\+ (J-K) \sum_{c,s} | \mathbb{N}_c\mathbb{N}_c \rangle \langle \mathbb{N}_s\mathbb{N}_s|.
\end{multline}

To study the symmetries of this Hamiltonian, we first consider
$\sum_c\ket{\mN_c\mN_c}$. Using an SU$(6)$ transformation $U$ on one
sublattice and $U^{*}$ for its complementary sublattice. This leads to
\be
\sum_c\ket{\mN_c\mN_c}=\sum_{a,b,c} U^a_c U^{*b}_c \ket{\mN_a\mN_b},
\ee
which reduces to $\sum_b\ket{\mN_b\mN_b}$ as $U$ is unitary, and thus preserves
the form. For terms such as
$\sum_{c}\ket{\mN_c\mN_{\cp}}\bra{\mN_{\cp}\mN_{c}}$, we transform
using $U$ on both sublattices, leading to a preservation of the form using
similar arguments. The above statements imply that for a Hamiltonian with
both terms invariant, we would require $U^{*}=U$. This condition is satisfied
by elements of the orthogonal group SO$(6)$, which comprises of real matrices
which generate proper rotations in six dimensions.

We note that an almost identical transformation is found in Ref.\cite{keselman2020dimer} for a SU$(4)$
antiferromagnet and in Ref.\cite{kaul2012spin} for a spin-1 biquadratic model
on the triangular lattice.

\section{Energy, order parameters and their Binder cumulants near the phase transition}\label{AppBC}

In this section we present a detailed description of numerical data near the quantum phase transition located at $\theta_c \simeq -0.5969(1)\pi$ for both the energy and Binder cumulants of order parameters.

\begin{figure*}[t]
\includegraphics[width=0.9\linewidth]{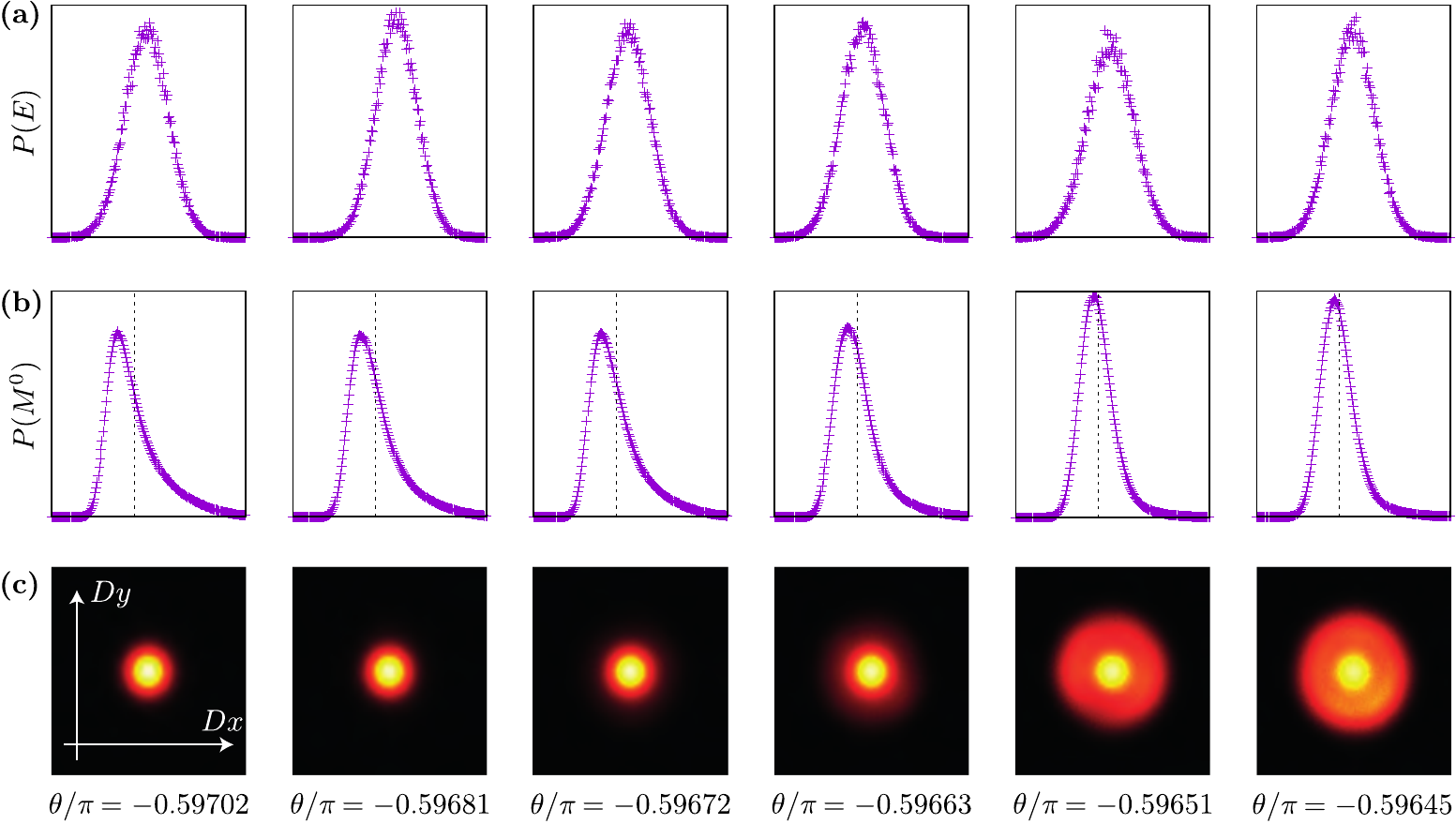}
\caption{Histograms for energy (a), nematic along with a vertical line
marking the mean (b) and VBC (c)
for $L=32, \beta=2L$ and values of $\theta$ on either side of the phase transition.}
\label{fhist32}
\end{figure*}

{\it Energy histograms --- } A first order phase transition can be detected, if strong enough, by the existence of two peaks in the histogram of energy (recorded during the Monte Carlo simulations) corresponding to energies of the two coexisting phases. In the top panel of Fig.~\ref{fhist32}, we present energy histograms for a system size $L=32$ for different values of $\theta$ close to and across the quantum phase transition, where we observe no sign of such double-peak feature.

{\it Nematic order parameter distribution --- }
For the SO$(6)$ version of the Hamiltonian, each site can take one of 6 colors.
To study the nematic ordering we use a 6-dimensional nematic order parameter
as defined in the main text.
In the disordered phase, $M^c$ is expected to have a Gaussian
distribution with mean zero and independent of all $M^{c'\neq c}$.
This implies that a Binder cumulant defined as
$U_{M_c}=\frac{\braket{(M^c)^4}}{\braket{(M^c)^2}^2}$
evaluates to three in the disordered phase. In the ordered phase,
$U_{M_c}$ evaluates to a finite value which is not unity due to
the SO$(6)$ symmetry. This can be observed in the histograms of the nematic order parameter shown in
Fig.~\ref{fhist32} (middle panel), as crossing the quantum phase transition. The distribution changes from Gaussian in the
VBC phase where the nematic order parameter is disordered (right side of the panels), to
a skewed distribution whose shape is dictated by the underlying SO$(6)$
symmetry in the nematic phase (left side). Once again we find a lack of double-peak distributions, showing
consistency with a continuous phase transition on length scale $L=32$.

To understand the shape of
this distribution, consider first a sample product state drawn from the
Monte Carlo simulation in the nematic $\mN$ basis. Note that in this basis
the nematic phase corresponds to a simple SO$(6)$ ferromagnet.
Let us denote the fraction of sites hosting color $\mN_c$ as $a_c$.
As we expect nematic ordering, without loss of generality, $a_0$ can be assumed
to be larger than all other $a_c$, and all other $a_c$ equal due to the
remnant symmetry between the non-dominant colors. Now consider the operator
$M^0_i=\ket{\mN_0}\bra{\mN_0}$ acting at site $i$. Using the shorthand
$\ket{c}=\ket{\mN_c}$ only for this section, we see that $\braket{c|M^0_i|c'}=1$
for $c=c'=0$ and 0 otherwise. As the product state of the system is
representative of the ordering, we must include all states reached by SO$(6)$
rotations starting from this state. This can be engineered in a straightforward
manner by applying the rotation on $M^0_i$ using an SO$(6)$ rotation matrix $O$
as $O^TM^0_iO$. Due to the constraints on $M^0_i$, this reduces to the matrix
$A_{kl}=O_{0k}O_{0l}$. As we are working with a product state, applying this
at site $i$ in state $c$, we get $\braket{A}_i=(O_{0c})^2$. As we have assumed
that the fraction of sites in state $c$ is $a_c$, $\sum_i\braket{A}_i$ reduces
to $\sum_c a_c(O_{0c})^2$. Using the conditions that all $a_c$ are equal except
$a_0$ and $\sum_c a_c=1$, we can write $a_0=\frac{1}{6}+r$ and
$a_{c\neq 0}=\frac{1}{6}-\frac{r}{5}$. This implies that $\sum_c a_c(O_{0c})^2$
can be broken into $(\frac{1}{6}-\frac{r}{5})\sum_c(O_{0c})^2+
(r+\frac{r}{5})(O_{00})^2$. We can reduce the first term by using the
identity $OO^T=I$, which implies $\sum_c O_{0c}O^T_{c0}=\sum_c O_{0c}O_{0c}=1$.
This leaves a dependency on the SO$(6)$ matrix given only by $(O_{00})^2$,
which must be averaged uniformly over all realizations of the rotation matrix.
As $O_{00}$ is one component of a unit vector chosen at random in
six-dimensional space, its distribution can be calculated analytically
by considering a particular value of the first component. The probability
of this value lying between $x$ and $x+dx$ is given by the volume of the
five-dimensional shell over which the rest of the components are distributed.
Using the expression for the surface area of a five-dimensional sphere, we
can deduce that $p(x)\propto(1-x^2)^{3/2}$.

We use the above arguments to calculate the theoretical prediction for the
value of the Binder cumulant $U_{M_c}$ in the nematic phase. First, we note that
$M^0$ for $U_{M_0}$ is defined to have a zero mean, i.e,
$M^0=\sum_i (M_i^0-1/6)$. The relevant powers to be calculate for $U_{M_0}$
are $\braket{(M^0)^4}$ and $\braket{(M^0)^2}$. Let us first begin with
the quadratic term. Expanded in the site index, this assumes the form
$\sum_{i,j}(M_i^0-1/6)(M_j^0-1/6)$. Under an $SO(6)$ rotation, each term 
(denoted by $A_{ij}$ for convenience) in
the sum transforms to $[O_{ak}(\delta_{a0}-1/6)O_{al}]_i
[O_{bm}(\delta_{b0}-1/6)O_{bn}]_j$, where repeated indices are summed over
and $[]_a$ indicates that the operator acts on site $a$.
Now we can evaluate $A_{ij}$
in the product state where the state at site $i(j)$ is given by $c_{i(j)}$.
This leads to $\braket{A_{ij}}=(O_{0c_i}^2-1/6)(O_{0c_j}^2-1/6)$.
The double sum over all sites, $\sum_{ij}A_{ij}$, can now be written in a
factorized form as
$\sum_{c}a_c(O_{0c}^2-1/6)\sum_{d}a_d(O_{0d}^2-1/6)$.
Each individual sum in this expression has already been evaluated to
$(r+\frac{r}{5})(O_{00})^2-\frac{r}{5}$. Using the probability distribution
of $x=O_{00}$ discussed in the paragraph above, we can now express
$\braket{(M^0)^2}$ as the integral $(1/N)\int_{-1}^1 (6x^2-1)^2p(x)dx$,
where $N$ is the normalization of the probability distribution, given by
$\int_{-1}^1 p(x)dx$. A similar analysis for the fourth power leads to
$\braket{(M^0)^4}=(1/N)\int_{-1}^1 (6x^2-1)^4p(x)dx$. Combining these
results, we can conclude that the value of Binder cumulant in a nematic
ordered state is

\begin{equation}
U_{M_0}=\frac{\frac{1}{N}\int_{-1}^1 (6x^2-1)^4p(x)dx}
{\big[\frac{1}{N}\int_{-1}^1 (6x^2-1)^2p(x)dx\big]^2}=\frac{114}{25}.
\end{equation}

We find that
the expectation
$U_{M_0}=\frac{114}{25}=4.56$.
is in agreement with the Monte Carlo
simulations presented below in the region of parameter space where
we expect nematic ordering.

{\it VBC order parameter distribution --- }
To detect VBC ordering, we use $D^2=D_x^2+D_y^2$ (with $D_x=\sum_i (-1)^{i_x} C_{i_x,i_y}\cdot C_{i_x+1,i_y}$ and $D_y=\sum_i (-1)^{i_y} C_{i_x,i_y}\cdot C_{i_x,i_y+1}$ as in the main text) and similarly define the Binder cumulant
as $U_D=\frac{\braket{(D^2)^2}}{\braket{D^2}^2}$. In the disordered phase
$(D_x,D_y)$ form a two-dimensional Gaussian distribution leading to $U_D=2$.
In the ordered phase, $U_D=1$ as fluctuations in $D^2$ are small compared to
its mean value. Note that $D^2$ is sensitive only to the development of
non-zero VBC ordering and does not differentiate between various types of VBC
orderings, such as columnar and plaquette.

The histograms for the VBC order parameter shown in the bottom panel of Fig.~\ref{fhist32} all show a circular shape but with a finite radius that decreases as one moves towards the nematic phase (the finite value of the left panels located in the nematic phase are associated to the finite size $L=32$).

\begin{figure}[t]
	\begin{minipage}{\columnwidth}
		\centering
		\includegraphics[width=\linewidth]{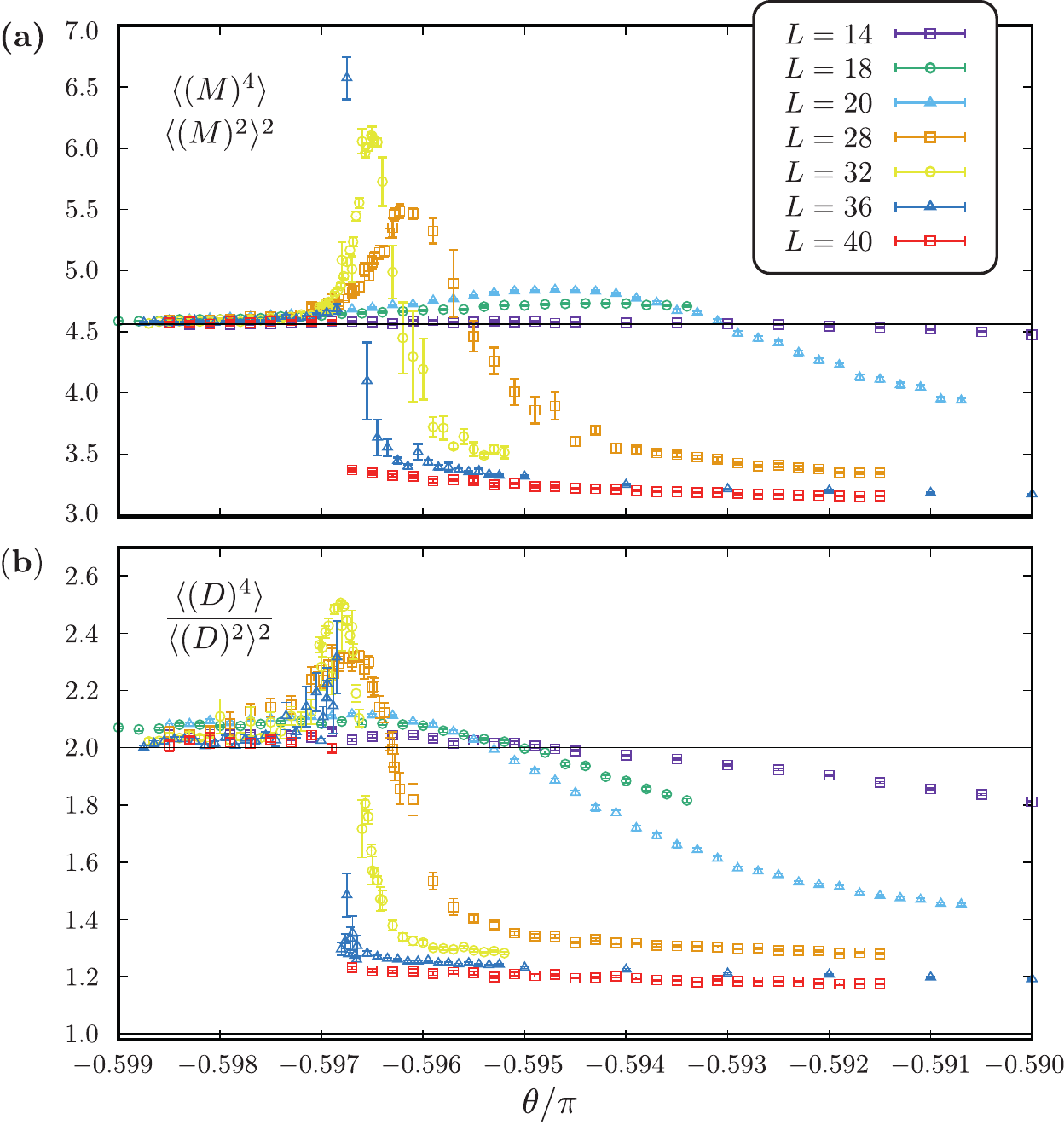}
	\end{minipage}
	\caption{(a) Binder cumulant of the nematic order parameter and
		(b) the VBC order parameter showing a single critical point, and strong
		non-monotonic behavior.}
	\label{fbcum}
\end{figure}

{\it Binder cumulants --- }
We finally present in Fig.~\ref{fbcum} the values of Binder cumulants as a function of $\theta$ close the phase transition, for different system sizes.
We observe a non-trivial non-monotonous behavior for both nematic $U_{M_c}$ (top panel), and VBC $U_D$ (bottom panel) Binder cumulants.

For the nematic Binder cumulant, data on small systems range within the disordered value $3$ (reached for large enough $\theta$) and the expected ordered value $4.56$ (reached for $\theta < \theta_c$). On the other hand, starting from $L\simeq 18$, the Binder cumulant curve overshoots the {\it ordered} value as one approaches the transition point $\theta_c$ from above, with curves showing a steeper overshoot as $L$ is increased. For a first order transition, a somewhat similar behavior is predicted~\cite{vollmayr_finite_1993} on the basis of a two-peak distribution of the order parameter (which we do not observe, see above) resulting in a value of the Binder cumulant at the maximum scaling with volume $L^2$.
We have checked that the maximum of $U_{M_c}$ does not scale as the volume $L^2$, at least on the lattice sizes accessible to us. Curves for different system sizes cross at different values of $\theta$, which is usually indicative  of a first order transition (but note however the very narrow range of $\theta$ displayed in Fig.~\ref{fbcum}). The non-monotonous behavior does not allow to conclude on the order of the phase transition (in particular a data collapse is not satisfying), but we note that the sharp overshoot feature is converging towards our estimate of $\theta_c \simeq -0.5969 \pi$ obtained from stiffness crossing (see main text).

A similar, albeit slightly different, non-monotonous behavior is observed for the VBC Binder cumulant, with a somewhat smoother overshoot over the {\it disordered} value of the Binder cumulant. Here again the maximum does not scale with volume, and could actually be converging to a finite value given the data on the largest systems that we could simulate ($L=36, 40$). The maximum anomaly also converges towards our estimate of $\theta_c$.

Overall we conclude that the Binder cumulants of both order parameters do not display the behaviors expected either at a continuous phase transition (no clear unique crossing point) or at a (strong) first-order phase transition (with an anomaly scaling as the volume of the system size).

\section{Nature of the U($1$) symmetry in the VBC phase}\label{appC}

\begin{figure}[t]
	\begin{minipage}{\columnwidth}
	\centering
	\includegraphics[width=\linewidth]{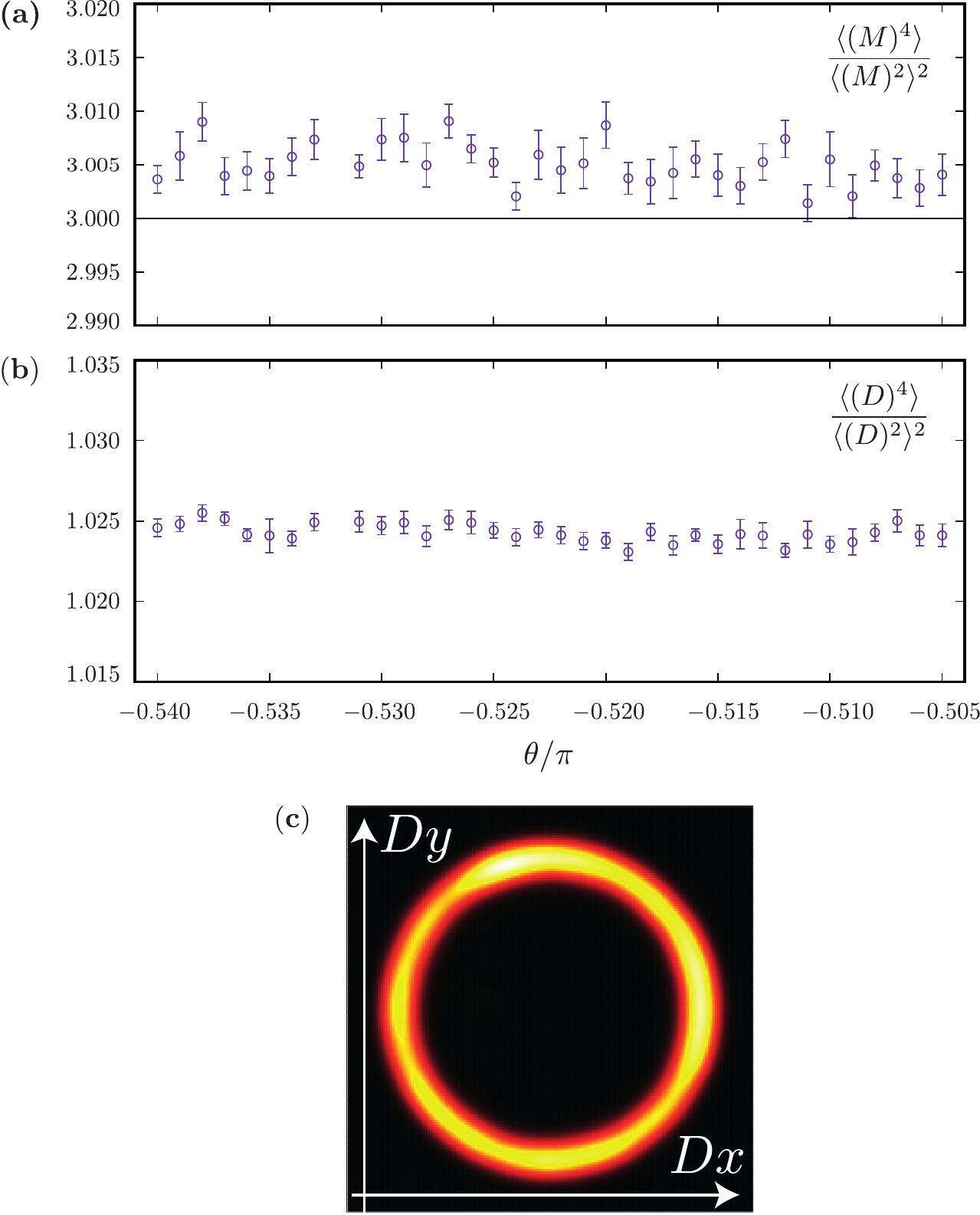}
\end{minipage}
	\caption{Binder cumulants for nematic (a) and VBC (b) order parameters
		for $L=96$ and $\beta=24$. (c) Heat map shows histogram for the
		VBC order parameter for $L=96,\theta=-0.52\pi,\beta=12$.}
	\label{fbcum96}
\end{figure}

Here we show evidence for the nature of the VBC phase by studying a large $L=96$
lattice. In order to determine the nature of the phase, we display in the top panel of Fig.~\ref{fbcum96}, the nematic Binder cumulant in the range $[-0.54 \pi, -0.5\pi]$ and we clearly see that it approaches close to the
expected value of $3$ in the disordered phase. On the other hand, the VBC Binder
cumulant (middle panel is close to $1$) in the same  range, as expected for an ordered VBC state.

\begin{figure}[!h]
	\begin{minipage}{\columnwidth}
	\centering
	\includegraphics[width=\linewidth]{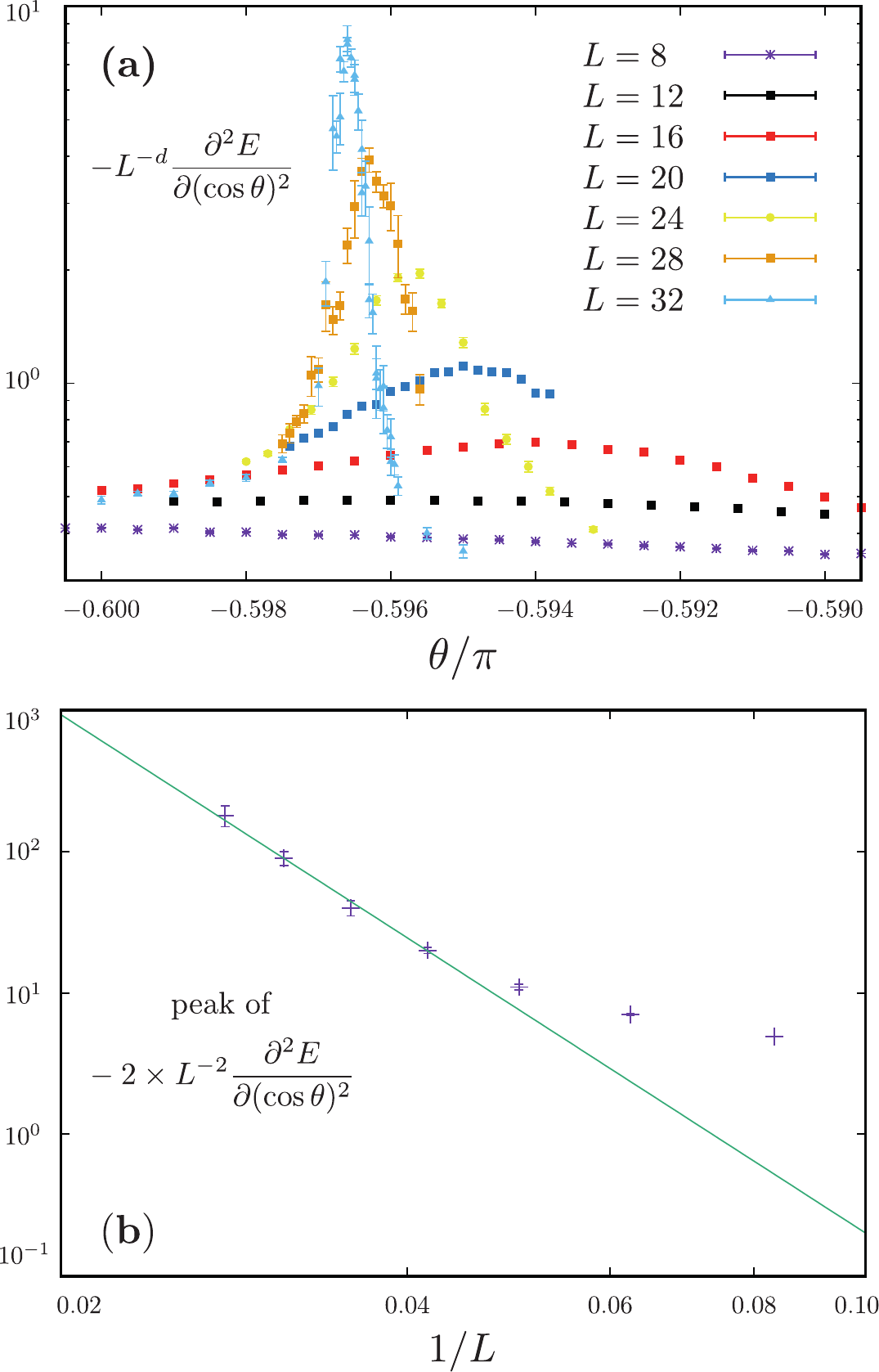}
\end{minipage}
\caption{(a) The negative second derivative of the ground state energy
per site
$-L^{-d}\frac{\partial^2E}{\partial(\cos\theta)^2}$ calculated using $\beta=2L$
shows a divergence close to the phase transition at $\theta=-0.596 \pi$.
(b) Scaling of the peak with system size on a log-log plot, fit to
$aL^b$ yields $b=5.2(3)$.}
\label{fenergyd2}
\end{figure}

This preliminary check being performed, we now seek for the specific symmetry breaking pattern of the VBC. We find that
even at this large system size $L=96$, there is no obvious discrete symmetry breaking, as we report in the bottom panel of Fig.~\ref{fbcum96} for $\theta=-0.52 \pi$.
There the sample histogram of the VBC order
parameter at a relatively low temperature of $\beta=12$ clearly displays a U($1$) symmetry.
Note that we are unable to simulate lower temperatures for $L=96$ due to ergodicity constraints and finite statistics of our simulations, as the system is able to sample only a portion of, and not the full, circle. As seen for the Binder cumulant of the VBC order
parameter in Fig.~\ref{fbcum}, this finite statistics issue does not affect the estimation of the magnitude fluctuations.
Recalling that a component of $\vec{M}$ is defined as
$M^s=\frac{1}{N}(\sum_i\ket{\mN_s}\bra{\mN_s})-\frac{1}{6}$, we see that
$\langle(M^s)^2\rangle$ yields a non-zero value for nematic ordering.

\section{2nd derivative of energy}

The second derivative of the ground state energy per unit site w.r.t
$\cos(\theta)$ can be calculated using the formalism developed in
Ref.~\cite{albuquerque2010quantum},
where the Hamiltonian is of the form $H_0+gH_1$, and the derivative is
calculated w.r.t $g$. Since our Hamiltonian is of the form
$\alpha H_1+\gamma H_2$ with $\alpha=\cos\theta$ and
$\gamma=\sin\theta-\cos\theta$, we have to consider the
derivative for both terms and the expression reduces to

\begin{multline*}
-\frac{\partial^2E}{\partial(\cos\theta)^2}=
\frac{1}{\beta}\bigg[\frac{A_1}{\alpha^2} +\bigg(\frac{1}{\gamma}\frac{\partial^2\gamma}{\partial\alpha^2}+\frac{1}{\gamma^2}\bigg(\frac{\partial\gamma}{\partial\alpha}\bigg)^2\bigg)A_2\\
+\bigg(\frac{2}{\alpha\gamma}\frac{\partial\gamma}{\partial\alpha}+\frac{1}{\gamma}\frac{\partial^2\gamma}{\partial\alpha^2}\bigg)A_{12}\bigg]
\end{multline*}
with
\begin{align*}
&A_1=\langle N_1^2\rangle-\langle N_1\rangle-\langle N_1\rangle^2, \\
&A_2=\langle N_2^2\rangle-\langle N_2\rangle-\langle N_2\rangle^2, \\
&A_{12}=\langle N_1N_2\rangle-\langle N_1\rangle\langle N_2\rangle.
\end{align*}
where $N_{1(2)}$ corresponds to the number of operators of type $H^{1(2)}$
in an operator string generated by the stochastic series expansion and $\langle \ldots \rangle$ the standard Monte Carlo average.

The negative second derivative estimated using the expression above is displayed  around the expected phase transition in the top panel of Fig.~\ref{fenergyd2}.
We observe that it diverges with system size, with a maximum approaching the critical point.

At a continuous quantum  phase transition in dimension $d$, the second derivative of the energy is expected to scale as $-L^d \frac{\partial^2 E}{\partial  (\cos(\theta))^2} \propto L^{2/\nu - (d+z)}$ where $\nu$ is the correlation length exponent and $z$ the dynamical critical exponent~\cite{albuquerque2010quantum}. Assuming a continuous phase transition takes place and that $z=1$ (see scaling of the stiffness in the main text), a fit of the divergence of the peak (shown in the bottom panel of Fig.~\ref{fenergyd2}) leads to an exponent $1/\nu=4.1(2)$, which is anomalously quite large.

We conclude that while a divergence of the second derivative of the energy is compatible within system sizes $L\leq 32$ with a continuous transition, the anomalously large value of the effective correlation length exponent that we obtain $1/\nu=4.1(2)$ hints towards a first-order character of the phase transition, which is confirmed by the time trace presented for larger system size in the main manuscript.

\section{Comparison with SO(5) nematic to VBC transition}\label{appD}

\begin{figure*}[t]
\includegraphics[width=0.9\linewidth]{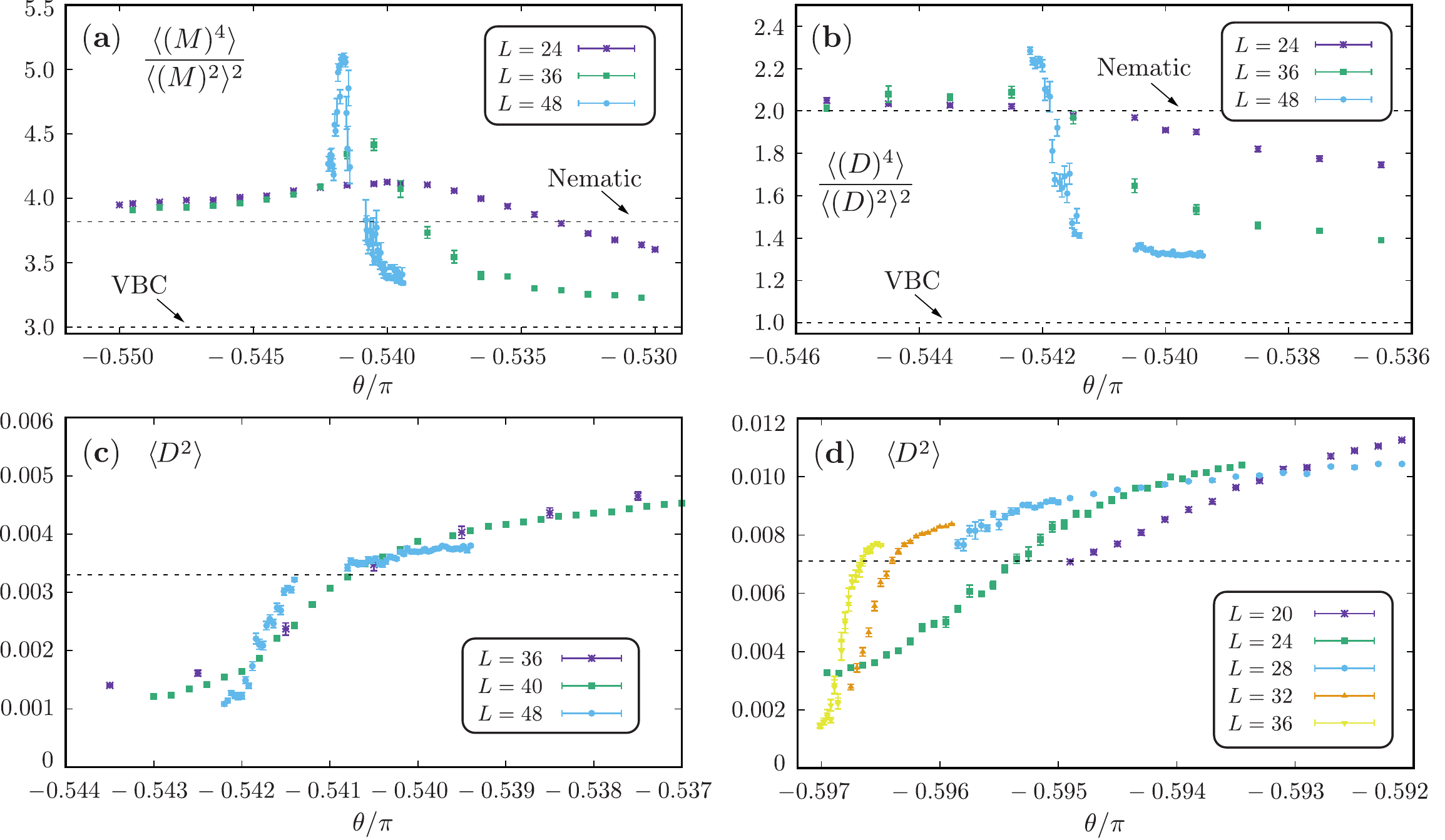}
\caption{
Binder cumulants for nematic (a) and VBC (b) order parameters for $SO(5)$
symmetric microscopic degrees of freedom as a function of
$\theta/\pi$, where constant lines mark the expected values in the two phases.
(c) and (d) show the expectation value of the VBC order parameter 
for $SO(5)$ and $SO(6)$ respectively. A crossing point is visible with increasing
system size, and a rough estimate of the thermodynamic value of the
discontinuity in the order parameter is given by dashed constant lines.
}
\label{Fig5_supp}
\end{figure*}

To understand the change in the nature of the transition with changing
number of components accessible to the microscopic nematic degree of freedom,
we simulate the nematic Hamiltonian (Eq.~\eqref{eqHnem}) for 5 possible
colors on each site. These simulations are motivated by the relevance of SO(5) symmetry for {\it e.g.} spin-3/2 fermionic cold atom systems~\cite{Wu2003,Wu2006}. 

 In the phase space region defined by 
$\theta/\pi\in(-0.75,-0.5)$, we find a nematic and VBC phase, separated by
a direct transition, similar to the $SO(6)$ case studied in the main
text.  The behavior of both Binder cumulants is shown as a function of
$\theta/\pi$ in Figs.~\ref{Fig5_supp} (a) and (b).

To identify the nematic phase, we calculate the Binder cumulant
of the nematic order parameter defined similarly as in the $SO(6)$ case. Repeating the argument above for the case of an $SO(5)$ symmetry, we find $p(x)\propto (1-x^2)$ for the distribution of the first component and that the Binder cumulant is re-expressed
as $\displaystyle U_{M_0}=\frac{\frac{1}{N}\int_{-1}^1 (5x^2-1)^4p(x)dx}
{\big[\frac{1}{N}\int_{-1}^1 (5x^2-1)^2p(x)dx\big]^2}=\frac{42}{11}
$. 
 
We find (Fig.~\ref{Fig5_supp} (a)) that the nematic Binder cumulant tends to the predicted theoretical value $42/11 \simeq 3.818$ in the
parameter range $\theta/\pi\in(-0.75,-0.5443(2))$, beyond which we find
a VBC phase, indicated by the approach of the VBC Binder Cumulant to unity (Fig.~\ref{Fig5_supp} (b))
We also observe the development of a non-monotonic
behavior with increasing size similar to the
$SO(6)$ case, indicating a possible first order
transition.  

While it is difficult to differentiate between weak and very weak first order phase transitions given the large scale lengths involved and the large number of components in these models, we now present two numerical observations which lead us to
conclude that the first order nature for $SO(5)$ is weaker than the same
for $SO(6)$.

The first of this is the ergodicity achieved by our QMC algorithm for
sizes close to $L=48$ for $SO(5)$. As we have shown in the main text,
the algorithm suffers from strong metastability for a size of $L=40$ for $SO(6)$, making
it impossible for us to get reliable data for larger sizes. This feature is
absent for $SO(5)$ at least till sizes of $L=72$. This shows that the
transition is not of a strong first order nature, where we would expect the
algorithm to oscillate between two qualitatively different phases.

The second observation involves the behavior of the VBC order parameter
close to the transition as it approaches zero. Both
$SO(5)$ (Figs.~\ref{Fig5_supp} (c)) and $SO(6)$ (Figs.~\ref{Fig5_supp} (d)) show crossing points in the VBC order parameter, which
are not expected at a conventional continuous transition.
This allows us
to estimate the size of the discontinuity in the VBC order parameter at
the transition (assuming that it is first order)
and we show a rough
estimation of the thermodynamic discontinuity in both plots using dashed
constant lines. A comparison of Figs.~\ref{Fig5_supp} (c)
and (d) shows that the discontinuity for $SO(6)$ is roughly a factor of 2
greater than that for $SO(5)$, also suggesting that the $SO(5)$
symmetry realises a weaker first order transition.

\end{document}